# Table of Contents

# Abstract

Renewable energy systems are an increasingly popular way to generate electricity around the world. As wind and solar technologies gradually begin to supplant the use of fossil fuels as preferred means of energy production, new challenges are emerging which are unique to the experience of decentralized power generation. One such challenge is the development of effective monitoring technologies to relay diagnostic information from remote energy systems to data analysis centers. The ability to easily obtain, synthesize, and evaluate data pertaining to the behavior of a potentially vast number of individual power sources is of critical importance to the maintainability of the next generation of intelligent grid infrastructure. However, the application space of remote monitoring extends well beyond this.

This paper details the development and implementation of an open-source monitoring framework for remote solar energy systems. The necessity for such a framework to be open is much better understood when considered through the lens of the theoretical potential for remote monitoring technologies in developing countries. The United States and other industrialized nations in the so-called 'first world' are likely to be slow to seriously adopt renewable energy on account of the massive investment and infrastructural changes required for its integration into the existing electrical grid. In countries where grid infrastructure is generally inadequate or nonexistent, this barrier is far less of a concern, and renewable energy technologies are viewed more as an enabling tool for progress than as a disruptive and expensive technological tangent. In this context as well, remote monitoring has a role to play.

Of course, renewable energy systems are nothing new in the developing world—they are just inaccessibly expensive to most individuals. Still, international charity organizations have been integrating renewable energy technologies such as solar power systems into their development projects for more than 40 years. In Sub-Saharan Africa in particular, many of these efforts have resulted in failure. Chief among the culprits responsible for these failures are the implementing organizations themselves, who almost pathologically fail to transfer the knowledge required to maintain renewable energy systems to local stakeholders. While a pervasive lack of access to technical training in most developing countries does not bode well for the success of future endeavors to promote electrification—rural or urban—it is arguable that remote monitoring systems would be of nontrivial assistance in such efforts. Of course, cost is still the greatest barrier to entry with respect to any given technology in the developing world. Therefore, *this* project revolves entirely around the use of open platforms and inexpensive, generic technologies to produce a viable remote monitoring framework to be used in environments where the resources and general knowledge required to maintain renewable energy systems is particularly scarce.

# 1. Introduction

Remote monitoring is not a new or unsolved problem. Therefore, the rationalization for pursuing this particular project requires some explanation and a description of the context in which the necessity for open-source remote monitoring arises. Furthermore, some vocabulary needs to be modified for specificity's sake. Renewable energy systems encompass a vast range of technologies, but in particular this project is concerned with solar technology. Additionally, the results of this project, while arguably generalizable to any given solar application anywhere in the world, are specifically intended to demonstrate how remote monitoring can be made useful and affordable in developing countries. Finally, so as not to make callous generalizations about the homogeneity of the so-called 'developing world', the requirements for this project have been designed with the realities of poverty and underdevelopment specific to Sub-Saharan Africa in mind. The applicability of the results of this project in a different region or environment is left for others to decide.

*1.1 Road Map*

In order to argue the need for open-source alternatives in solar remote monitoring, it is important to demonstrate the unsatisfactory nature of the status quo. To do this, certain premises need to be established, in order.

1.) By some mechanism, photovoltaic, or solar, technology has become a sufficiently common alternative to the use of grid-powered electricity in Sub-Saharan Africa, enough to the point where generalizations about its use can be made.
2.) In many situations in Sub-Saharan Africa where solar power systems are implemented, these systems quickly fail as a result of poor maintenance.
3.) While the proper maintenance of solar power systems requires regular to intermittent attention on the part of trained individuals, a significant portion of the maintenance process is an information technology problem.
4.) An information technology solution in the form of a remote monitoring system can serve a critical role in providing system overseers with the information required to maintain solar power systems.
5.) Existing solar remote monitoring systems are expensive, limited in their application, and for the most part proprietary.
6.) Such a system should be available as an open-source technology because of the economic realities of poverty and underdevelopment in Sub-Saharan Africa.

Arguing for these premises will establish the proper context within which to discuss the requirements for this project and evaluate the results of this semester's attempt at such an implementation. Thereafter, future implementations, abstractions, and applications can be discussed in a productive way. The overall purpose of this thesis, moreso than a description of an implementation or a celebration of achievement, is to serve as a proof of concept that solar remote monitoring is neither expensive nor particularly cumbersome to implement and thus warrants further investigation and development by the open source community. There are many applications for a system like this. Monitoring a solar power array is just one of the possibilities. Practically any device with measurable outputs running in a remote environment represents a potential future extension of this project. The hope, of course, is that others will be able to build upon this framework and use the results described here to cultivate their own applications. Advancements in this field can yield cheaper and more robust solutions to assist in both the maintenance and viability of remote solar power systems.

# 2. Background and Literature Review

The first logical premise to establish in the defense of this thesis is the assertion that solar technology is a more or less commonplace method of alternative energy production in Sub-Saharan Africa. The second is that many solar energy systems fail as a result of misuse. Because these two premises are generally part of the same narrative when discussed in the context of development in Sub-Saharan Africa, they will be established together in this section through an in-depth analysis of three separate case studies. To begin, however, it deserves to be stated that when we consider the notion of solar power as common we are discussing the subset of people who actually have access to electricity, and thus generalizations drawn from studies of the utilization of solar energy systems are capable of only limited abstraction.

## 2.1 Solar Energy and International Development

In 2002, the African Energy Research Policy Network (AFREPREN) published an article in Energy Policy magazine which estimated that roughly 68% of the inhabitants of Sub-Saharan Africa live in rural areas without access to grid-powered electricity. In the conventional interest of development and poverty reduction, the question of how to provide modern energy services to this enormous proportion of the population is of critical importance. As many African governments have proven incapable or unwilling to tackle this issue, the general consensus of the international development community at large has been to emphasize the dissemination of renewable energy technologies to rural areas. This focus on localized power generation has in turn led to the implementation of a variety of developmental programs involving the use of solar technology to provide generally inaccessible communities with electrical power for important infrastructure like schools and hospitals.

Now, short of making the problematic assertion that electricity is a solution to poverty, suffice it to say that the status quo has prompted many charity and non-governmental organizations (NGOs) to use solar energy systems in conjunction with their humanitarian development efforts. With this context in mind, arguments can be framed about best practices for sustainable development and the role of remote monitoring in renewable energy projects.

## 2.2 Solar Energy Case Studies—Popularity and Practicality

How has solar technology fared as a solution to the pervasive and systemic lack of access to electricity in Sub-Saharan Africa? This is not a question which can be disputed on account of the maturity of the renewable energy approach. Rural development projects in Sub-Saharan Africa involving the use of photovoltaics for electricity generation have been underway since the 1960's, and, therefore, the accumulated body of literature on this subject is vast. Upon examination of this literature, the most striking observation to consider is the general lack of theoretical disagreement among case studies. This is not to say that the experience of every solar project or private initiative has been similar; rather, that they all seem to succeed or fail for similar reasons. In rudimentary terms, there has been a great diversity of contextual experiences with solar technology. In Zimbabwe, for example, the United Nations Development Program Global Environment Facility (UNDP-GEF) project from 1993-1997 seems to be the reigning example of overall failure, while the Nyimba Energy Service Company (ESCO) project in Zambia in 2000 and the Namibian government's ongoing 'Home Power!' program appear to be shining arguments for the continued proliferation of solar technology.

Overall, these projects were both introduced into comparably poor rural communities, however, the character of each project's implementation was markedly different, and the end results appear to reflect this. The experience of the UNDP-GEF project seems to have left the reputation of solar technology permanently scarred in some circles. Nevertheless, solar technology remains popular. In some ways, it appears that independent of the direction of the community of non-governmental organizations solar technology continues to find its way into the hands of those who can afford it. There are plenty of viable alternatives to solar technology as far as energy generation goes, and therefore the phenomenon of solar technology's continued popularity is of particular interest to analyze and explain.

*2.2.1 Solar is popular…*

Ray Holland, a member of the Intermediate Technology Development Group (ITDG), argued in an issue of IEEE Review in 1989 that the cost of solar technology needs to fall considerably before it can be used for applications in developing countries beyond communications, lighting, and water pumping. Today, 20 years later, this is still the primary barrier to the increased dissemination of solar technology. In spite of this, however, solar technology remains a popularly sought after and highly demanded commodity in Sub-Saharan Africa. The reason for this may be explained in part by the observation Holland makes that the over-consumption of electricity is generally not a problem for rural communities. The fact that solar technology can provide high-quality lighting and allows for the use of radios, small television sets, and cellular phones is enough to fuel the continued demand for panels and batteries. Mark Hankins and Robert J van der Plas, in their research for the World Bank's Energy Sector Management Assistance Program (ESMAP) in Kenya in 1998, concluded that between 75% and 90% of rural Kenyans *know* about solar technology. While not conclusive, they argue that the continued popularity of solar technology may be attributable to just that—its popularity. Absent readily available information about other sources of renewable energy such as wind turbines, efficient biomass combustion, and micro-hydroelectric generators (which are argued by some to be better suited for rural energy needs), it is probable that rural dwellers aspire to own solar panels because it is a symbol of relative social status and represents a step out of poverty.

*2.2.2 …But should it be?*

Two researchers for AFREPREN, Stephen Karekezi and Waeni Kithyoma, conducted an evaluative study of renewable energy strategies for rural African communities in 2002. Their main criticism of contemporary approaches to energy generation on the part of international development organizations is actually the gross *over*-emphasis of solar technology. Karekezi and Kithyoma are keen to point out that solar systems are woefully inaccessible to the vast majority of rural communities. Citing the failure of previous micro-finance and subsidy-driven solar distribution programs, it is consistently estimated that around 80% of the rural poor in Sub-Saharan Africa cannot afford even the smallest 18W solar power systems, let alone keep up with service and maintenance fees. Putting relative costs into perspective, for a 40-50W solar power system, it is estimated that most rural households would have to pay on average 200% of their per capita GNP just to afford start-up and installation costs. In the United States, this percentage would amount to an average cost of $50,000 for the same system. Additionally, Karekezi and Kithyoma argue that home-based energy needs in Sub-Saharan Africa are 90% to 100% comprised of cooking and heating. As solar technology is generally limited to lighting and communication, it is hardly a viable or worthwhile investment for most rural families to purchase even the smallest solar power system.

However, Karekezi and Kithyoma do admit that solar technology can be employed to provide quality lighting and offset the need for fuel-burning light sources. In fact, solar is actually preferable to biomass in terms of lighting. Biomass applications produce low-quality light and require the continued purchase of fuels for combustion, whereas solar technology combined with high efficiency CFLs can perform for many years with relatively little maintenance. This also goes to directly offset a large portion of the exposure to smoke particulates created by combustion lighting, (which is an altogether different public health crisis that is beyond the scope of this paper). It is not Karekezi and Kithyoma's purpose in their analysis to completely discredit the application of solar technology, as they do recognize the usefulness of the services it can provide insofar as lighting, entertainment and communication are concerned. They do, however, stress that solar is quite limited in its application and should not be at the forefront of renewable energy strategies for rural development.

*2.2.3 The UNDP-GEF Project*

The analysis of Karekezi and Kithyoma is important as an African perspective on current energy paradigms. While it is clear they are opposed to the dominance of solar technology in rural development strategies, it is important to consider the context from which they draw their criticisms. One of the major case studies that exemplifies the failure of development projects involving solar technologies is the UNDP-GEF program in Zimbabwe from 1993-1997. The GEF project was one of the largest efforts by the UNDP in history to proliferate the use of solar technology, and the fact that its results are so widely criticized deserves examination. In a May 2000 article of *Energy Policy* magazine, Tim Jackson, Tinashe Nhete, and Yacob Mulugetta published a comprehensive article on the lessons of the GEF project. Jackson and Mulugetta are researchers from the University of Surrey Center for Environmental Strategy (UK), and Nhete is an ITDG member from Harare, Zimbabwe.

The main criticisms that Jackson, Nhete, and Mulugetta cite, overall, are the lack of rural stakeholders in the projects' success, and the lack of post-project follow-up by the UNDP. In a nutshell, the project was a donor-driven program to install 10,000 environmentally friendly solar power systems in Zimbabwe's rural areas over the five-year period between 1993 and 1997. From the very outset it can be argued that the project was overly ambitious and attempted to address the concerns of too many incompatible interests. In the official report on the project released by the UNDP in May 2004, the mission statement includes the achievement of the UN Millennium Development Goals, the satisfaction of environmentalist concerns regarding greenhouse gas emissions, the alleviation of rural poverty in Zimbabwe, and the creation of new markets for local solar companies. In the attempt to satisfy all of these goals at once, the GEF project spread itself very thin and failed to do much beyond meeting donor funding deadlines. In the *Energy Policy* article, the GEF goal of reducing global carbon emissions by installing environmentally friendly technologies in rural areas of Zimbabwe is correctly likened to "using a sledgehammer to crack a nut." It is an ostensibly ridiculous assumption to say that rural communities in Sub-Saharan Africa have an even measurable impact on global carbon emissions, and it is an act of blatant hypocrisy to impose upon them such strict limitations as a prerequisite for development aid.

*2.2.4 Critical Implementation Failures*

Despite the wrong-headedness of the GEF project from the outset, critical failures were made during the project's implementation which undermined the likelihood that the systems put in place would remain functional for much longer than the project itself. While some of the solar power systems were donated to rural communities, the majority of them were sold through micro-financing schemes. In this respect, the most common problem facing the proliferation of solar technology was again brought to bear. Only 20% of rural households in Zimbabwe could afford even the smallest solar system offered to them by the GEF project. In this respect, the locally affluent became the primary benefactors of the project rather than the rural poor. Moreover, emphasis on spurring private sector solar industries was given priority over the transmission of knowledge to the owners of solar power systems. In turn, local solar technology suppliers and businesses were the only resource rural communities had when they experienced system failures as a result of misuse.

On its head, this is not an entirely unworkable situation. Technical knowledge of solar system design and implementation is a marketable skill and should be allowed to seek professional outlets. The problem arises, however, when the owners of solar power systems have *no* understanding of how their systems work and must pay local technicians for even the most minor causes of concern. Over time, it was found that a majority of people misused their systems without realizing why, and were then burdened by the increased costs of maintenance and repair.

*2.2.5 The Problem of Maintenance: Battery Care*

The most common problem was the routine overuse of batteries. Solar power system owners were not educated on the basic principles of battery care, and were not aware of their system's limitations. Thus, they routinely left electrical loads on until their battery banks fully discharged and their appliances cut off. (Solving *this* problem in particular is one of the primary goals of the remote monitoring system described in this thesis.) Then, without knowing the consequences of consistently over-discharging batteries, this practice was repeated until the battery electrolyte was depleted and could no longer hold charge. While this is the eventual fate of any rechargeable battery, understanding safe discharge limits can double or triple battery longevity. At the very least, if solar power system owners were simply taught that they should only leave their lights and appliances on for a certain amount of time so as to prevent their batteries from completely discharging before allowing them to recharge, it is likely that many of the UNDP-GEF systems would have lasted much longer. (This may be overly optimistic, as revealed by the results of the next case study.) The way the UNDP-GEF project was conducted and the overemphasis placed on the role of local technicians led to many systems failing far before they should have. Many solar power system owners then replaced their failed batteries with cheaper and more affordable car batteries, which, unfortunately, are not designed for the charge cycling of a solar power system, have fewer amp-hours, and, in turn, would end up failing soon after. While a robust understanding of the physics involved in this process is useful, it is not necessary to solve such problems. Much of the failure of the UNDP-GEF project to create lasting improvements in rural communities is hinged on the fact that they did not transmit even the most basic knowledge of solar power system ownership to the project's benefactors.

*2.2.6 The Problem of Maintenance: Local Technical Support*

Part of the reason the importance of a basic understanding of solar power system maintenance and care seems to have been overlooked by the UNDP-GEF project was that it was hoped this void would be filled by the growth of local businesses and technicians. In the interest of time, perhaps, this was wishful thinking on the part of the UNDP-GEF project planning staff. It also appears that another casualty of the UNDP-GEF project's donor-imposed time constraints was the formation of a stakeholder community. No local or international NGOs, rural authorities, or patrons of any sort were procured prior to the full fledged implementation of the project, much to the dismay of observers in Zimbabwe and elsewhere. The UNDP-GEF project, it seems, was constrained so tightly by its five-year commitment to install 10,000 solar power systems that it forgot most everything else and left the responsibility of repairs, maintenance and education up to unproven and—more importantly—*undesignated* local actors.

*2.2.7 The Pitfalls of Scope*

Another unfortunate complication of the UNDP-GEF project was that its immense scope had the unintended consequence of undermining the long term viability of local solar technology businesses. The introduction of the raw equipment for over 10,000 solar power systems into the local market dramatically distorted the prices of system components. The parallel market which formed in the midst of the UNDP-GEF installations also took a toll on the ability of registered solar technology businesses to function. Cheaply made amorphous silicon panels from South Africa made their way into Zimbabwe and began to compete with the more expensive multi-crystalline silicon panels supplied by the UNDP. Panel theft also became a problem. Solar panels are valuable commodities and are easily removed from rooftops as a result of the necessity that they are open and exposed. The primary targets of theft were women, the elderly, and the disabled. The resultant fear of criminals and the loss of such a large investment made some potential buyers of solar power systems turn the opportunity down. This, combined with the onslaught of economic downturn in Zimbabwe, drove many of the newly formed solar power companies out of business in a relatively short period of time. In 1997, there were roughly 60 registered solar companies to service 10,000 new customers as a result of the UNDP-GEF project. By 2000, there remained only 15. Of the businesses that survived, the vast majority of them had been in business prior to the UNDP-GEF project. Today, with inflation rates in Zimbabwe above 1 million percent, it is doubtful that even the strongest of these solar businesses still exists.

*2.2.8 The Lack of Follow-Up*

Unfortunately, there is no post-project data on either the UNDP-GEF project systems or the businesses it tried to create. The reason for this, while hardly surprising given the circumstances, was that the UNDP had not planned on making any post-project assessments and instead assumed that this data would be collected by local solar energy companies. As over 75% of the businesses created by the UNDP-GEF project failed within three years, there is today no data on the performance of *any* of the 10,000 installed systems.

The experience of the UNDP-GEF project tarnished much of the popular support among some NGOs for similar endeavors. However, it deserves to be stated that the UNDP-GEF project was poorly executed and failed in a rather predictable manner. The tiger's share of the blame in this instance can be laid at the feet of the UNDP for their almost inspired incompetence and their treatment of communities of interest in the project moreso as a commodity in service of an environmentalist publicity stunt than as the intended benefactors of a serious and rigorously researched effort at promoting sustainable rural electrification. Fortunately, while the UNDP-GEF project may be sadly representative of the general experience of NGO projects involving solar energy systems, other projects have been much more successful, despite being faced by similar challenges.

*2.2.9 Glimmers of Hope: The Nyimba ESCO Project*

The failure of the UNDP-GEF project is humbling, but nevertheless deserves to be countered with examples from smaller, but more effective development efforts involving solar energy systems. One such example of a successful solar energy program was implemented by the Nyimba Energy Service Company (ESCO) in Nyimba, Zambia in 2000. In this case, the introduction of solar technology had a positive impact on the lives of people in rural communities, particularly with regards to educational prospects. Mathias Gustavsson and Anders Ellegaard, two researchers from Goteborg University in Sweden, reviewed the progress of this ESCO project in a 2004 article of *Renewable Energy* magazine.

An 'ESCO project' is a new type of technology-dissemination program which has been employed in a variety of contexts and communities around the world with considerable success. The Nyimba ESCO project consisted of 100 individual solar home systems installed in rural communities near the town of Nyimba. The general philosophy of the Nyimba ESCO project is that expensive technology such as a solar power system is beyond the purchasing capability of most rural dwellers and should therefore be given as a service package, whereby clients enter into a contract providing them with the installation of a 50W solar home system in exchange for a monthly service fee of 25,000 Zambian Kwacha, or the equivalent of U.S. $6.85. This service fee covers any problems that may arise during the time clients use their solar home system, including the replacement of parts if they are damaged. The contractual system was composed of a 96 Ah deep cycle battery and charger, a 12V, 50W solar panel, and four 7W CFLs for lighting, including fixtures. As the combined wattage of the CFLs is only 28W, clients were encouraged to buy their own appliances to make use of the rest of their solar home system's capacity. This was meant to endow clients with a sense of ownership in the maintenance of their solar home system; of course, most of the need for technical expertise was intended to be accounted for through monthly servicing fees.

*2.2.10 Poverty and Servicing Fees*

One of the unique facets of the Nyimba ESCO project compared to other case studies from Sub-Saharan Africa is the fact that people who would not otherwise be able to afford a solar home system were given the opportunity to use one for a small monthly fee. In practice, however, this fee was still a barrier for many households. The results of the Gustavsson-Ellegaard study found that 90% of the households with solar home systems contained at least one formally employed person. This division in incomes between formally and informally employed households was further underscored by the fact that between 10% and 15% of households with solar home systems found it extremely difficult to pay their fees. Nevertheless, it was also found that around 50% of clients did attempt to expand and maximize the use of their solar home systems, some of whom even managed to purchase and install inverters to run an appliances requiring alternating current (AC) power. Overall, however, most households acquiring a solar home system felt that the primary benefit of solar technology is the availability of quality lighting at night.

*2.2.11 The Benefits of Solar Technology*

The benefits of quality light manifested themselves quite vividly qua the experience of the Nyimba ESCO project. The Gustavsson-Ellegaard study found that nearly 60% of clients claimed they could not read at night prior to having a solar home system, and 50% believed that children were the primary benefactors.

Around 89% of households with a solar home system claimed that their children used the available light at night to study, whereas only 42% of households without a solar home system could claim that their children attempt to study at night. Interestingly, it was found that children would study together at night in houses with solar home systems. Furthermore, teachers began to use the advantage of having dependable light at night to teach classes. Extremely poor children who cannot afford schooling and must work to support their families during the day were able to benefit from classes taught at night. This was also found to be the case in Namibia (the next case study), and highlights the educational promise that solar powered lighting may indirectly hold for rural communities in Sub-Saharan Africa.

The ability to expand one's active day was cited as another benefit of having a solar home system. Twenty percent of businesses owners claimed they could expand their working hours after dark with the aid of dependable lighting. Beyond this, the desire to own a television set was shown in the Gustavsson-Ellegaard study to be greater than the desire to have a solar-powered water pump. This seems to point to a consistently exhibited desire on the part of rural communities to have access to appliances and commodities that allow for entertainment and increased communication with the outside world. As one teacher in Nyimba was paraphrased as saying, "Our lifestyle changes; it is like we moved from the rural area to the town. We now have light in the evenings and we can play music."

*2.2.12 The Recurring Problem: Battery Care*

Still, amid this apparent success in Zambia, some of the pitfalls of the UNDP-GEF project were also found to exist. A general lack of knowledge among clients absent the support of local technicians seemed to lead to the overuse and failure of batteries. The Gustavsson-Ellegaard study revealed that even with the regular monthly support of local technicians that 25-30% of the installed battery banks failed after only 2 years of use. The life-span of an average deep-cycle battery is in the range of 5-8 years. It is thought that if systems were operated a bit more carefully that this life-span could be increased, but the underlying point remains that lack of training on the proper care of battery banks is one of the chief reasons that solar power systems fail when introduced into rural environments.

*2.2.13 Glimmers of Hope: The Namibian Home Power! Program*

Throughout this discussion of the various experiences with solar energy in Sub-Saharan Africa, it deserves reiteration that development projects in this context tend to succeed or fail for similar reasons. In situations where the owners of solar power systems are trained in the use of battery banks and/or local technicians are specifically *designated* as system overseers, these systems can be of great and lasting benefit to rural communities. In situations where this does not happen, they rather quickly fail. To further establish this point, another successful solar energy program worth mentioning is the Namibian governmental 'Home Power!' program.

Njeri Wamukonya, a researcher for the United Nations Environmental Program Collaborating Centre on Energy and Environment (UNEPCCEE) in Denmark, prepared an evaluation of the Namibian government's post-independence efforts to gradually expand the electrical grid to all parts of the country. To date this has been an enormous endeavor, and thus the Namibian government has launched a low interest rate loan program called Home Power! through which rural and semi-rural communities can purchase home solar power systems. The program provides applicants with a solar home system installation to be paid back over a maximum of five years at a 5% interest rate.

The experience of Namibia has been similar to that of the rest of Sub-Saharan Africa in the sense that, again, only a minority of rural households seem to be able to afford even the smallest PV systems, and therefore localized elites tend to gain more from the dissemination of solar technology than do the rural poor. In fact, the Home Power! program does not grant installations to applicants who do not already make enough money to afford the systems.

*2.2.14 Stakeholder Cultivation and Knowledge Transfer*

The Namibian government has committed itself to the development of its grid infrastructure and popular electrification in a way that NGO and donor-led programs simply cannot sustain for any considerable length of time. Furthermore, the Home Power! program, in contrast to the UNDP-GEF debacle, is a long term effort and does not have deadlines to install a specific number of solar power systems in random rural communities. Home Power! involves the contracting of local suppliers to install systems properly in client's houses and emphasizes the transfer of knowledge regarding maintenance and installation between technicians and clients. This is a responsible action on the part of the Namibian government to attempt to prevent its installments from being wasted on account of misuse. This is a critical point: The Namibian government has drawn from the general corpus of knowledge on the performance of solar power systems in Sub-Saharan Africa, and their policy *reflects* the fact that training clients on the maintenance of their systems, regardless of their technical background, is one of the most crucial aspects of a successful solar energy dissemination program. There are national radio programs, advertisements, and TV commercials in Namibia intended to educate citizens on the proper use and limitations of their solar energy systems. This kindles local businesses, encourages proper usage of solar power systems, and promotes the technology around the country.

An interesting note about the Namibian Home Power! program is that recipients of solar home systems, by virtue of the fact that they understand exactly what their system can and cannot be used for, report almost universally that their welfare improves after installation. (Interestingly, households with off-grid power do not have to deal with blackouts and actually have more consistent lighting than do their counterparts in urban areas!) The primary benefits of quality lighting from solar technology were reported as the ability to read at night, listen to radio, and in some instances watch television. Above solar water pumping or any other appliance, surveys indicated, again, that the first things people with solar arrays desire to run are television sets. This allows people to watch news and keep updated on national affairs, watch sports, and in general provides a greater sense of connectedness with the outside world. Such additions to rural people's daily lives, while primarily aesthetic in nature, are widely reported as improvements over life without electricity at all.

## 2.3 Lessons Learned From Case Studies

The case studies listed here have been chosen for their exemplar nature in demonstrating the realities of the utilization of solar energy in Sub-Saharan Africa. (There are many more accounts to consider, and the curious or perhaps unconvinced reader is encouraged to peruse the References section.) Of course, some analysis is required to crystallize and properly abstract the lessons of these experiences. An obvious question that arises from the case studies is why solar energy, given its costs and limitations, makes any sense in the first place as a developmental motif in Sub-Saharan Africa? This is a valid question, but not one to be explored in any critical depth here. Perhaps unsatisfyingly, this project does not attempt to derive an 'ought' about the use of solar energy from the 'is' of its use by development organizations and governments as a way to promote rural electrification. The subsequent premise that remote monitoring can help in the maintenance of solar power systems is established only as a statement of fact. This thesis makes no attempt in any serious way to assert a preference that solar energy 'ought' to be used in Sub-Saharan Africa or that this specific project 'ought' to be employed, but rather that if the proper functioning of such technologies is considered desirable, then this is one viable solution.

Moving on, the case studies discussed here are sufficient to demonstrate the first two premises of the justification for this project, that 1.) Solar technology is a commonly used alternative to conventional grid-powered electricity and 2.) That many solar power systems fail as a result of misuse. However, the evidence from these case studies also establishes the first half of the third premise, that trained individuals are necessary for the proper maintenance of solar power systems. This is a fairly trivial point, but one that is made all the more poignant by the positive experience of the Namibian Home Power! program. When both the owners of solar power systems and technicians in the local community are actively involved in the maintenance process, systems last longer and perform better. In retrospect, the obvious nature of this fact does not seem to have occurred to the UNDP during the GEF project. (Why not?) Unfortunately, this is unsurprising in the broader context of international development. Humanitarian and charity organizations have long been the subjects of scorn in academic circles for their pathological ignorance about some of the most obvious truths about sustainable development.

*2.3.1 Sustainability in Theory and Practice*

Sustainability is an ironic topic in the field of humanitarian development projects because it is so often emphasized in theory but almost always bungled in practice. Consider the espoused benefits of renewable energy systems versus the practicality of their application: As has been shown, right off the bat, solar energy falls flat on its face as a practical investment for poor and rural communities because it is astronomically expensive given the services it can provide and the relatively meager economic returns it can yield through income-generating activities. In most situations it is simply a material impossibility to afford. Thus, in the vast majority of cases where solar technology actually finds its way into the hands of otherwise underdeveloped and impoverished communities in Sub-Saharan Africa, it is only through the work of governments and international charity organizations. As a result of this dependency, the sustainability of development projects involving solar power systems is tenuous. Solar energy may provide a theoretical source of clean, renewable, and essentially 'free' electricity, but its initial cost is so high that it usually necessitates external intervention.

Moreover, one of the major problems with the introduction of expensive and potentially complicated technology into underdeveloped communities is the concept of cost after installation. Once a new piece of infrastructure is in place, a permanent maintenance cost has also been introduced. If a solar array is installed on the roof of a school, a technician will have to routinely look after it, and that technician's time and skill-set cost money. But how can an already impoverished community afford such expenses? And where does the knowledge come from?

In the case of the UNDP-GEF project, this entire aspect of the long-term sustainability was ignored. The rest is history. In the case of the ESCO project in Zambia, while the same criticisms can be leveled with respect to the failure of some systems in the hands of owners without the proper training to maintain them, the fact that a community of locally designated technicians was established had a definite impact on the project's success. Of course, the scope of these projects may limit how much can be abstracted from them. The UNDP-GEF project involved 10,000 individual systems, whereas the Zambian project only involved 100 individual systems, and the Namibian project is ongoing. In general, of course, it is almost always the case that proper training translates to better maintenance.

Financing schemes are a more difficult issue to resolve. The programs in Zambia and Namibia were both instituted under the assumption that the initial costs of installation, (absorbed by the implementing organizations at the outset) would be eventually be repaid by the recipients of solar energy systems. The UNDP-GEF project sold the majority of its solar power systems through micro-finance schemes, and donated the rest. It is likely that an important tenet of promoting stakeholdership is a realistic concept of cost on the part of project benefactors. Still, it is clear that after-installation costs must as low as possible. Realistically there is almost no up-front or after-installation cost that will be low enough for the poorest rural families. But on this note, we are again faced with the inherently unsustainable notion of external dependency. Conventionally, when development organizations try to fill the gap in knowledge and resources alluded to above, they do so in unsustainable ways, or—as was the case for the UNDP-GEF project—not at all.

*2.3.2 The Conventional Approaches*

Some common approaches to nurturing development projects after their inception include the establishment of effectively permanent international fund-raising schemes (such as the approach taken by World Vision) or the periodic sending of NGO personnel to maintain a given project themselves. Such strategies are problematic because in general they tend to indicate a failure to thoroughly consider how best to *sustainably* empower communities of interest. They are also expensive, neglect the role of community stakeholders, and essentially beg the whole development question in the first place. At some point, the long-term viability of a given project needs to be critically examined. What lasting and truly sustainable development has actually occurred if, for instance, the proper functioning of a solar power system requires a life-long maintenance commitment on the part of its original installer? An appropriate analogy could be that so-called 'sustainable development' projects which are not designed or, in practice, able to subsist without continuous external intervention are the moral equivalent of a business trying to stay afloat by purchasing its own inventory.

*2.3.3 The Bottom Line*

In unfortunately common fashion, humanitarian development projects involving solar technology fail. The specific reasons for failure vary, but it tends to be the case that development organizations neglect to plan for the problems of cost after installation. The knowledge required to sustain a given project is not properly transferred, and community stakeholders are generally left by the wayside. (Unrealistic notions of reliability may also be partly to blame: Solar energy systems are often naively expected to run without maintenance.) As a result, these systems are unintentionally misused, their batteries are overdrawn to the point of depletion, and they eventually stop functioning altogether.

Situations like this are all too common in the aftermath of short-sighted development projects, and must be avoided if such efforts are to be taken seriously. A sustainable project is more than the hip photo-op aesthetics of a solar array in a small African town. True sustainability must consider the economic impact of the introduction of technological infrastructure, and have sustainable financing schemes built in to account for both the costs of maintenance and the training of community stakeholders. Anything less is a waste of time and money.

## 2.4 The Role of Remote Monitoring

The use of such an extensive argument to establish the point that solar technology is emphasized by development organizations and in many cases poorly implemented in practice may seem somewhat extraneous. Additionally, the concepts of sustainability and sustainable development in general might also appear at first glance to be off-topic when considering the core subject matter of an open-source remote monitoring system. In reality this entire context is essential, because the solutions to be proposed here are arguably incoherent without it. A remote monitoring system is *part* of the sustainability of a given project involving solar technology, precisely because it can be of such crucial assistance in the maintenance process (a point which will be explored in greater detail later). The viability of development projects is a function of their sustainability. The fact that the history of humanitarian development is one of mostly failure and disappointment is a crucial point to digest in the process of understanding this system. It is impossible to phrase a solution in such a way that properly addresses this context without first understanding the full scope of the issues. Furthermore, the realistic potential of these sorts of projects is often overstated. A sober and well informed approach serves to mitigate this tendency. This project does not purport to be a silver-bullet to solve the issues of solar power system maintenance in Sub-Saharan Africa. It is simply an attempt to discover a small piece of the problem that *may* be addressable through the development of *some* technology—in this case a remote monitoring system.

*2.4.1 Maintenance and Information Technology*

Development is a difficult process to get rolling in truly sustainable fashion. While in theory, considering the pitfalls of external dependencies, a more grassroots approach may be the preferred strategy, in practice it often makes more sense to try and spur growth through an initial investment, external or otherwise. The experiences of the solar energy dissemination programs in Zambia and Namibia in fact argue this. Given the economic realities of poverty, the only way to leverage the potential of solar energy technologies in underdeveloped communities is through charitable investment. It's just too expensive otherwise. So, assume such investment exists—because it does. (This fact is agnostic with respect to whether we agree that it *should* be there.) The problem, then, is how to manage costs after installation. It is arguable that if the costs of maintenance can be accounted for in a sustainable way, then use of solar technology can be of enormous economic and social benefit through the income-generating activities it can enable, as well as the benefits it holds for things like education in rural communities.

Maintenance is a fairly well-defined task in the case of solar energy systems. The essential challenge is the proper cycling of batteries based on available sunlight and expected energy demands. As supported by the aforementioned case studies and other literature, the critical variable in the longevity of any solar power system is the degree to which the battery bank is properly cycled. This means that when the battery bank is drawn too low, turn off the load! Power down the inverter! All that this process really requires is the ability to keep track of system state. Any technologies that can obtain and analyze performance data are thus of enormous utility. This is a problem of information, because the easier that system diagnostics are to acquire and distribute the more reliable and robust the maintenance process becomes. This is the third premise for the justification of this project.

*2.4.2 Remote Monitoring*

Enter remote monitoring. The proper maintenance of a solar power system hinges on the degree to which system overseers can monitor the output and performance of the solar panels and the battery bank, and take appropriate action if and when problems occur. Thus, if something were in place to track system state and initiate alerts when necessary, system overseers would arguably have all the inputs they need to do their job. Consider for example the idea of a community training center in rural Namibia that operates at night using power generated from a solar array. When it is sufficiently dark, the lights are turned on, and whatever activities the center is being used for can commence. At some point in the night, the battery bank will likely be drawn down to the low voltage cutoff point beyond which the system should no longer be used. Ideally, a monitoring system would detect this problem, and initiate an alert in some form (a call or text message) to a local technician (or the staff at the center) who will then indicate to the appropriate parties that it is time to conclude the night's activities. The system is then turned off and the batteries are ready to be recharged the next day.

A sustained and responsible cycle of charging and discharging batteries in this fashion will allow a solar power system to be maximized in its longevity and utility to a given community. Furthermore, if some remote logging and analysis functionality has been implemented, system technicians can monitor the performance of the system over time and be able to judge when important adjustments or routine maintenance need to occur. Allowing the critical diagnostic information about a solar power system to be readily obtained and analyzed anywhere at any time would render the quintessential task of maintenance trivial.

This is the fourth premise of this thesis. Since the maintenance of a solar energy system is essentially a problem of information, remote monitoring can supply a solution. Of course, remote monitoring is fundamentally auxiliary—it is only a way to assist system overseers in their existing role and cannot possibly be expected to replace them. Nevertheless, the chances for the long-term successful operation of a solar power system can be greatly improved through the use of remote monitoring because of the flexibility, transparency, and the quality of information it can provide.

## 2.5 An Examination of Existing Remote Monitoring Systems

At this point, the fundamental argument for the use of remote monitoring in the context of development projects in Sub-Saharan Africa has been established. Unfortunately, this justification is not sufficient to explain why remote monitoring requires an open-source solution. There are existing remote monitoring systems on the market, so why re-implement this functionality?

The problem with existing solar monitoring systems is that they are limited in their application, are expensive, and in most cases require paying service fees to a third party above and beyond the cost of basic communication. In places where solar energy systems are employed as a replacement for a lack of electrical grid infrastructure, this cost may be extraordinary so as to lead to the abandonment of the idea of remote monitoring altogether (the more likely scenario being that it is never introduced in the first place). Thus, the reality of expensive and proprietary remote monitoring technologies holds hostage the viability of renewable energy systems in developing countries. In order to be practical, the cost of a monitoring system needs to be tailored to the economic reality of the environment in which it is implemented.

*2.5.1 How Remote Monitoring Works*

Effectively exposing the problems with existing solar monitoring systems requires backtracking to a more fundamental discussion of how these systems work in the first place. To begin, solar power systems tend to come in two flavors–remote and grid-tied. Remote systems are not connected to the electrical grid and serve as a primary power source. Grid-tied systems are integrated with the electrical grid and tend to serve as an auxiliary power source. The function of a remote system is to charge a battery bank that is tied to an inverter, which produces generally consumable alternating-current (AC) power. The function of a grid-tied system is to feed directly to an inverter and supply power while the sun is shining, deferring to the grid when it is cloudy or dark. The system proposed and implemented here is developed with a remote system in mind.

Remote solar power systems use devices called charge controllers to apply charging algorithms to banks of deep-cycle batteries. Controllers are essential because using solar panels to charge batteries is not a trivial task; a delicate balance must be struck between the need for batteries to be charged using well-defined and consistent charging cycles and the fact that the output of solar panels can be inconsistent and erratic depending on the weather. The primary function of a controller is to prevent the battery from being overcharged by the solar array. Nowadays most charge controllers are equipped with microprocessors that maintain historical data about the amount of power produced by a system. The diagram on the next page is helpful in understanding the high-level components involved in a common solar power system.

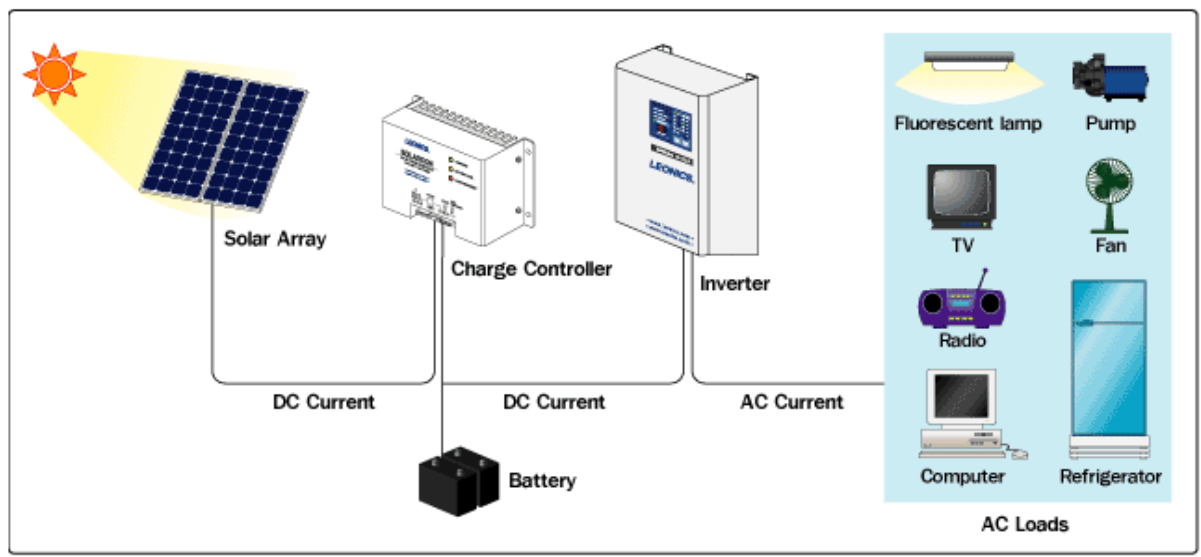

Fig 1: The Basic Design of a Solar Power System (Source: Leonics Co. LTD)

Monitoring is conducted by interrogating a charge controller about the performance of a solar power system, and then transmitting that data to a remote location or specific person. A cellular modem or other transmission device is generally the vehicle for this transmission. Many solar controllers have serial ports built into them that can be polled for data about a given system using some transmission protocol. Some controllers use open protocols to do this, others do not. A useful diagram to understand the high level structure of a remote monitoring system is shown below.

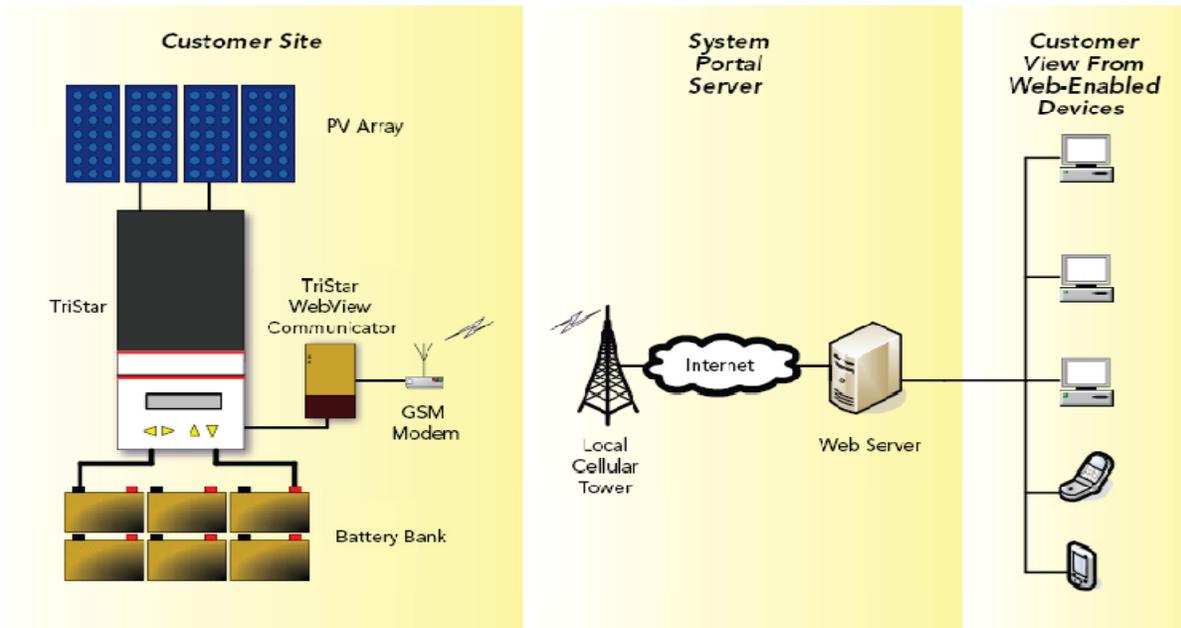

Fig 2: The Morningstar TriStar Web View remote monitoring system (Source: Morningstar Corporation)

For clarity's sake, the diagram in Fig 2 is of an actual industry product from Morningstar Solar. The 'TriStar' device is the solar controller. The next section discusses this and other systems in greater detail.

*2.5.2 Existing Monitoring Systems*

The industry leading solar monitoring system, according to Harald Kegelmann, CEO of Advanced Solar Technologies, Inc., a Gainesville-based solar energy installation company, is the Sunny Webbox, a remote monitoring device offered by SMA Solar Technology. SMA is a German solar energy equipment supplier. The Sunny Webbox itself is essentially a glorified modem that plugs into a grid-tied inverter and transmits data about a given solar power system via the Internet. SMA provides the hosting for the Sunny Webbox's web interface. Unfortunately, the communication protocol between the Sunny Webbox and an actual SMA inverter is not well documented or supported by SMA so it is difficult to program against, for the obvious reason that SMA prefers its customers to purchase SMA monitoring services. Furthermore, the Sunny Webbox requires wired Ethernet in order to remotely transmit data, which is an obviously crippling dependency if we intend to monitor devices operating in truly remote environments or in places without adequate internet infrastructure. Furthermore, the price tag of the Sunny Webbox device is on the order of $690 US. As it turns out, this is actually a fairly average price for such technologies.

Another industry-leading system is implemented by Fat Spaniel, a company that provides high quality web-based solar monitoring for a monthly fee (this is of course not including the hardware which must be purchased to interface with a solar controller). This service is slightly more robust than the service offered by SMA because Fat Spaniel employs cellular modems to remotely transmit data from a given solar power system. Furthermore, Fat Spaniel also recently began offering an open platform to expose their data to outside applications. Of course, this data is still only available for a monthly service charge of roughly $50. A range of services are available, and this fee can wind up being as low as $20 per month or as high as $120, depending on the options desired. One solar energy company that offers a monitoring service through Fat Spaniel is Morningstar Solar.

Morningstar Solar sets itself apart from SMA in one particular respect because they use the open Modicon Modbus transmission protocol for Programmable Logic Controller (PLC) devices to allow external applications to interrogate their controllers. They freely publish and provide the specification for their implementation of the protocol and thus allow other implementing technologies to be used with their controllers relatively easily. It is for this reason that this project is implemented using a Morningstar Solar TriStar-45 controller as an exemplar solar controller device. The Fat Spaniel hardware that is compatible with Morningstar controllers is listed at affordable-solar.com for the price of $1375. While Fat Spaniel services are more robust than SMAs, costs on this order of magnitude are an obvious deal-breaker for communities in Sub-Saharan Africa that might consider the use of remote monitoring to assist in the maintenance of their solar power systems.

Another monitoring service offered by Draker Laboratories, which specializes in commercial-scale remote monitoring services. While Draker services are of high quality and utility, their remote monitoring services are on a scale that is simply overkill for the vast majority of small solar home systems.

Outback Power is another solar energy technology company that produces excellent Maximum Power-Point Tracking (MPPT) controllers; however, they do not offer Internet-enabled remote monitoring services of any kind. This might not be an issue, necessarily, if an open protocol were defined to interrogate their controllers using a cellular modem or other such remote monitoring device. Unfortunately, the protocol used by their controllers to communicate with wired external monitoring devices is explicitly described in their documentation as proprietary.

This overview of some of the industry-leading solar controllers and monitoring technologies is intended to establish the fifth premise for the justification of the project; that existing remote monitoring systems are expensive, limited in their application, and require paying proprietary service fees to third parties. When viewed through the lens of the economic realities in Sub-Saharan Africa, even if remote monitoring hardware were provided by an international development organization as part of a solar energy dissemination program, even the most modest servicing fees and associated costs would likely be too great.

*2.5.3 How This Project Departs From Existing Monitoring Systems*

The final premise required for the justification of this project is established editorially. Existing monitoring systems do not meet or consider the needs of solar power applications in the developing world. This is not the fault of the corporations that implement them, of course. The reality is that remote monitoring has never been phrased as a problem in the context of the viability of solar power systems in rural communities in Sub-Saharan Africa. The contributing research to this thesis revealed no accounts of remote monitoring systems even being considered by development organizations undertaking solar energy dissemination projects. If a system were to be designed to work in this context, it would almost by definition have to be open-source. No proprietary servicing scheme through an international third party or major monitoring company could ever hope to be cheap enough. Furthermore, most of the remote monitoring systems on the market are overkill. The maintenance of a remote solar power system simply does not require such robust technology.

This project represents a significant departure from existing solar monitoring technologies because it is specifically intended to be open-source in every aspect outside of the actual solar power system itself. Currently there are no existing systems with similar intentions to what is being proposed by this project—indeed, if the string "open source solar remote monitoring" is typed into Google and searched, this project's development page is the first result.

At this point, this thesis has established the necessary premises to justify the implementation of an open-source monitoring system for remote solar power systems. The following sections are a description of one possible implementation.

# 3. Requirements and Implementation

If we accept the argument that maintaining a solar power system is essentially a challenge of obtaining critical information about the system and taking appropriate action, we can propose a solution in the form of a remote monitoring system. However, there are more concerns than just this. One of the greatest problems with remote monitoring, again, is its cost. A system that is intended to serve the purpose of remote monitoring in developing countries and rural environments must be inexpensive in addition to being effective. The most important factors to minimize are the costs after installation.

The up-front cost of a remote monitoring system is the hardware required to conduct the actual monitoring. After this cost is sunk, the problem becomes the cost of using the air waves to transmit data wirelessly. In the case of monitoring services, there is also a great deal of overhead built into servicing fees. This cost can be completely eliminated if a system is designed independently of a servicing company. While the expertise required for the maintenance of a solar power system is valuable, the expected service charges of any proprietary monitoring system are a deal-breaker in the context of developing countries. Of course, without resorting to illegal measures, it is practically impossible to transmit data over long distances wirelessly, so some cost has to be expected. Of course, this can be minimized, and this a topic that shall be discussed further in this report.

## 3.1 Solution Statement

To reiterate, the purpose of this project is to design and implement an open-source monitoring system for remote solar energy power systems that can deliver useful diagnostic information to system overseers.

This system is divided into two different parts. The first is a hardware-centric component that interfaces with a solar charge controller to determine diagnostic information about a power system. This component will process obtained data and transmit it via a telephony communication protocol to a server or, if necessary, a specific person.

The second component of the system is the recipient software on the other end of this transfer. This software shall store historical data about the solar power system and provide a web interface through which long term statistics can be determined and effective monitoring can be conducted.

*3.1.1 High Level Overview of the System: The Hardware Component*

The first part of this system is a hardware device that interfaces with a solar controller, polls it for information, and relays that information remotely to an appropriate party. This hardware component is designed with certain constraints in mind. First, it must not rely on customized hardware, or, if specific hardware must be utilized, it must be either open-source, or utilized in such a way that is agnostic with respect to any proprietary design detail. The solar controller itself is not considered to be a part of this specification because it can operate independently of any monitoring activity and is a general requirement of any functional solar power system.

*3.1.1.1 The Morningstar TS-45 Controller*

The solar controller that this project is designed to be compatible with (at least as a proof of concept) is the Morningstar TriStar-45 solar controller. This is on account of several factors, including personal experience working with Morningstar solar controllers (see Appendix D), the availability Morningstar controllers internationally, and the fact that applications can be easily developed to work with Morningstar controllers because they use the open Modbus hardware communication protocol.

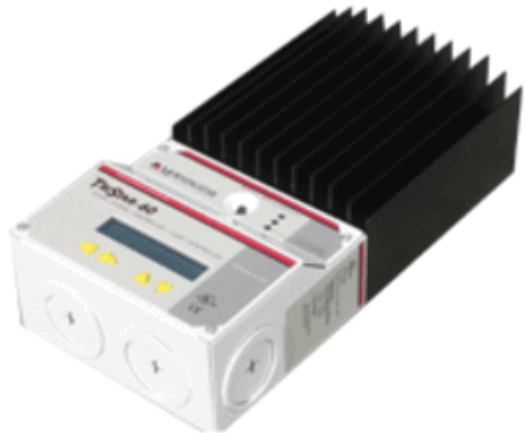

Fig 3: The TriStar-45 Solar Controller (Source: Morningstar Corporation)

The Modicon Modbus protocol reference guide is freely available online, and also includes sample code for more tedious implementation details such as the creation of a working Cyclic Redundancy Check (CRC) function. This specification was utilized to implement the software required to interrogate the TriStar-45 controller.

Morningstar offers all of the required documentation to develop applications in conjunction with their controllers online. The "TriStar Applications Guide", for instance, describes sample configurations for data acquisition and remote monitoring of solar power systems. This guide also unsurprisingly directs the reader towards further documentation regarding the monitoring service that Morningstar offers through Fat Spaniel.

The TriStar-45 has a serial RS-232 port built in which enables external devices to communicate with it. In a remote monitoring system, this is used as the interface between the data transmission hardware and the solar power system itself. The figure below shows the hardware interface of the TriStar-45. Item number 2 in this image is the RS-232 serial port.

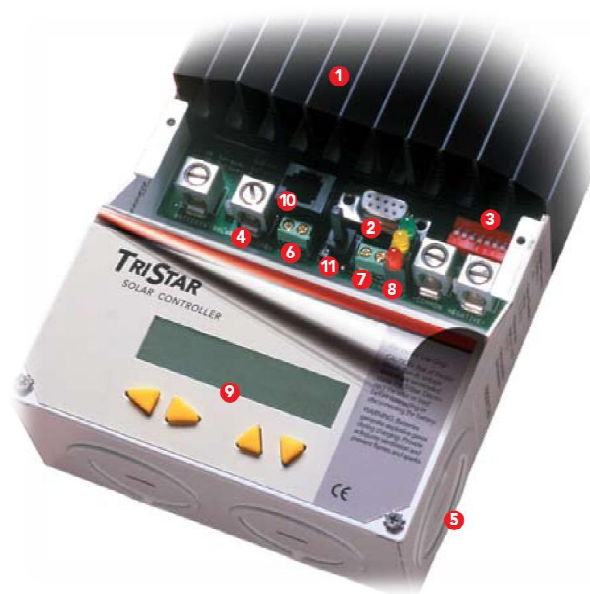

Fig 4: The Hardware Interface of the TriStar-45 (Source: Morningstar Corporation)

*3.1.1.2 Arduino and Freeduino Microcontrollers*

The other major technology that is utilized in this project is the Arduino electronics prototyping platform. Arduino is an open-source hardware platform intended mostly for hobbyists and casual hardware enthusiasts. Boards come in ranges of power and potential, and can be built by hand or purchased preassembled from organizations such as SparkFun. All hardware part listings, board schematics, and documentation required to build Arduino clones can be found and downloaded online. Guides can also be found to etch the Arduino PCB board by hand and solder it together. The software environment required to write and upload programs to Arduino boards is also freely available and completely open-source. Arduino boards and software are released under Creative Commons Attribution Share-Alike Licenses.

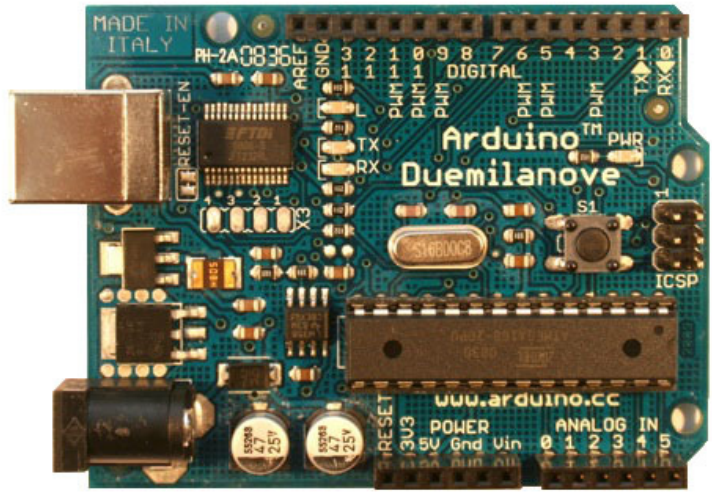

Fig 5: The Arduino Duemilanove (Source: Arduino)

The Arduino hardware environment is an ideal platform for a project like this because it allows the required hardware to be reproduced in any context where the parts can be found. The availability of parts is an obvious concern, but this is vastly preferable in the context of this project to purchasing a finished product from a specific company. In an effort to demonstrate the reproducibility of Arduino boards, this project opts to use a Freeduino board instead of an Arduino board. Freeduino is an open-source Arduino clone that can be purchased from hobbyist electronics websites as an unassembled bag of parts and a cut PCB board. (It also costs a fraction of the price of a preassembled Arduino board.) As part of the documentation for this project, an assembly guide for this board is provided online.

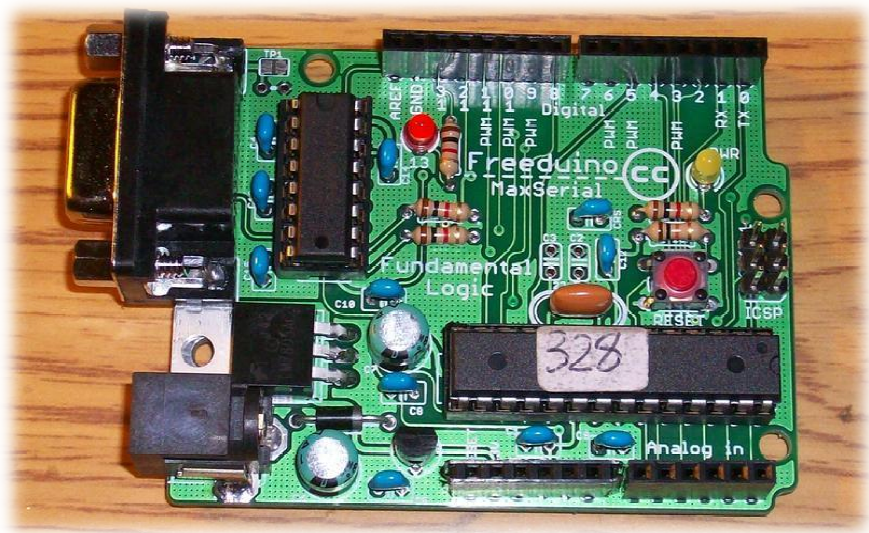

Fig 5: A Freeduino MaxSerial Arduino Clone

This project uses the Freeduino MaxSerial, which has an RS-232 port built in. This is required to interface with the Morningstar Solar controller, which allows external devices to interrogate it through an RS-232 9-pin serial port. (Most Arduino boards now use a USB port for serial communication, so the Freeduino MaxSerial is actually based on a now-obsolete Arduino specification.) The Freeduino board is based on the Arduino Decimilia specification, which uses an ATMega328 16 MHz microprocessor, and has 14 digital I/O pins, 5 analog input pins, and several voltage outputs for external circuitry.

*3.1.1.3 The Motorola W260g*

The last piece of hardware utilized in this project is a Motorola W260g cellular phone, which for the purposes of this project has been engineered into a data transmission device. This is an inexpensive disposable cell phone model generally sold by Tracfone, a pay-as-you-go cellular service company. In this project, the W260g serves as a replacement for an expensive cellular modem. As was the case in every example of existing remote monitoring systems, an expensive and overblown cellular modem is always employed as the actual data transmission device. Considering the fact that these devices are on the order of $600-$800, the exploration of the potential of disposable cellular phones is a worthwhile endeavor. The W260g is available for $14-$15 dollars. While the use of any given hardware generally couples a project to the nuances of a particular device, it is arguable that no part of the hardware design in this project is dependent on the specific design of the Motorola W260g. This will be explained in greater detail later.

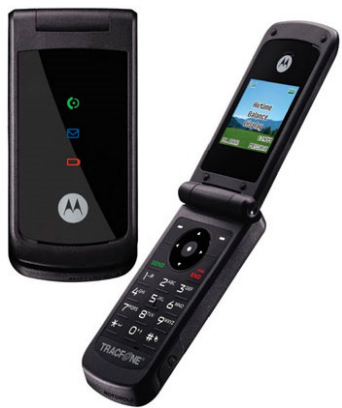

Fig 6: The Motorola W260g (Source: Amazon.com)

*3.1.1.3 How These Parts Fit Together*

The Freeduino MaxSerial is the centerpiece of the hardware component of this system. The MaxSerial uses the RS-232 serial port on the TriStar-45 to obtain critical diagnostic information about the performance of the solar panels and the batteries, and then manipulates a circuit built onto the motherboard of the cell phone to transmit that data over the airwaves as a string of DTMF tones.

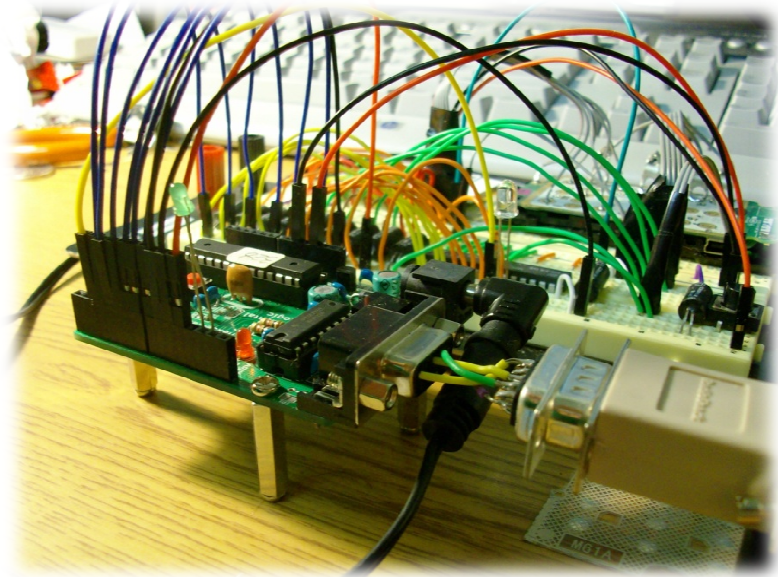

Fig 7: A Totally Cool Picture of the Hardware Component

*3.1.2 High Level Overview of the System: The Software Component*

The software system that interacts with and stores the data obtained from the hardware component is the second major part of this remote monitoring system. This software system is for most practical purposes independent of the hardware component. Its general purpose is to store historical data about a solar power system, and present it in such a way that allows for system analysis to be conducted. Such a system will accomplish this task by periodically calling the cellular phone of the monitoring device, collecting the data it receives in the form of DTMF, decoding the tones, and storing them in a database. This database can then be queried in order to produce graphs and other output representing such things as charge curves, solar insolation data, average power output over time, and so on. If a problem is detected, this system might opt to send an SMS message or other notification to one or many system overseers, but this notification has the obvious potential to be delayed heavily if a scheduler only prompts an update on the hour or every few hours.

*3.1.2.1 Requirements Vary…*

In reality the specification for this software system is fairly arbitrary. It could be a robust, extensible, and highly functional system, or it could be rudimentary and simple. The output format of the hardware system will always be the same, so it is entirely a matter of what the implementing party decides upon for requirements. The design of the hardware side of this monitoring system attempts to reduce the coupling between the hardware and software sides of this project as much as possible, but there is some agreement that must be instituted. For example, the fields in a data frame, the units of the numbers, the precision of the decimal places, and the order in which they are transmitted are things that cannot be easily changed. If a given software system required a different or more specific set of data parameters, the implementation of the hardware system would have to change, potentially significantly.

Still, knowing the order and being able to supply meaning to the content of a data frame received from a transmission does not give away the implementation details of the hardware in this project. This is simply the consequence of defining a new transfer protocol. If the protocol is open and well documented, it is easy for any implementing software application to use and build upon. The details of this project's protocol are explained in later sections.

## 3.2 Implementation and Design Considerations

Now, with a clear picture of the requirements and components involved with this system, the design considerations that went into this project's implementation can be explained in detail. This section is less intended as a tutorial on the construction of this project than it is a narrative of the dialectic of design and implementation.

### *3.2.1 Building the Hardware Component*

The hardware component of this project is the only thing that was truly completed during the course of this semester. While the implemented circuitry and the interactions between the Freeduino, the cell phone, and the TriStar-45 are not all that complex, the process at arriving at the finalized design was an extremely tumultuous road of wrong turns, mistakes, and frustration.

#### *3.2.1.1 Constructing the Freeduino*

Originally, there was an extensive period of research and inquiry into potential microcontrollers that could be used for the purpose of polling the TriStar-45 and transmitting its information. One of the first ideas was to use an already designed cellular-enabled development board with a GSM chip; however, the only company that really sells programmable chips that could be used for this purpose is Telit. This violates the constraint that this project not be designed with specific hardware in mind—that is, unless it is completely open source. Arduino boards turned out to be the ideal solution to this problem, again, because of the freely available schematics and open-source programming environment. Online resources for Arduino applications are also vast because of the size of the hobbyist community. They are also extremely cheap.

Of course, in order to explore the viability of reproducing and physically constructing an Arduino board (as would almost certainly be the case if this system were implemented in Sub-Saharan Africa), it became apparent that the best way to go would be to use a Freeduino board, purchased as an assortment of parts. Additionally, the latest Arduino boards are incompatible with the serial interface of the TriStar-45 because it uses a USB as its primary serial I/O port. While pins 0 and 1 on the Arduino have been made available for doing UART serial communication, the fact that the Freeduino MaxSerial board has a built in RS-232 port made it an ideal candidate for the job. In an effort to uncover as many complications that might arise from the construction of such a board, the online resources for this project include a detailed part listing and construction guide.

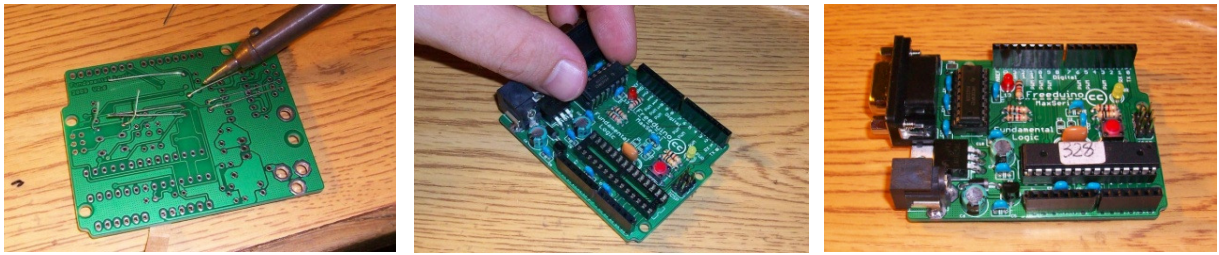
Fig 8: Building the Freeduino Board

One lesson in particular that emerged from the process of constructing the Freeduino board is the fact that any implementing party must be skilled at soldering and likely needs to have a basic understanding of such rudimentary concepts as the polarity of diodes and electrolytic caps. This is a potential complication in an environment where such skills are hard to come by.

*3.2.1.2 Constructing the Phone Circuitry*

The next issue was the use of a disposable cellular phone as a transmitter. Just how should such a device be utilized, and what are its constraints and limitations? This was not a subject matter for which many resources were readily available. The first item of inquiry was whether the microprocessor of the cell phone could be manipulated directly and used to transmit information.

An observation of the complicated machine-soldered micro-circuitry involved with any common cell phone quickly put this idea to rest. The next logical question involved the identification of ways to manipulate the phone with a microcontroller. The best option turned out to be the creation of some external circuitry to directly manipulate the keypad. This would enable calls to be made and received, and for data to be transmitted over a connected phone line via the DTMF keypad. The issue, then, was how to create this external circuitry. After taking one cell phone apart, it was quickly discovered that pressing the keys on the keypad is a fairly straightforward process of depressing a metal cap on a circuit board that in turn connects two concentric copper leads on the cellular motherboard together.

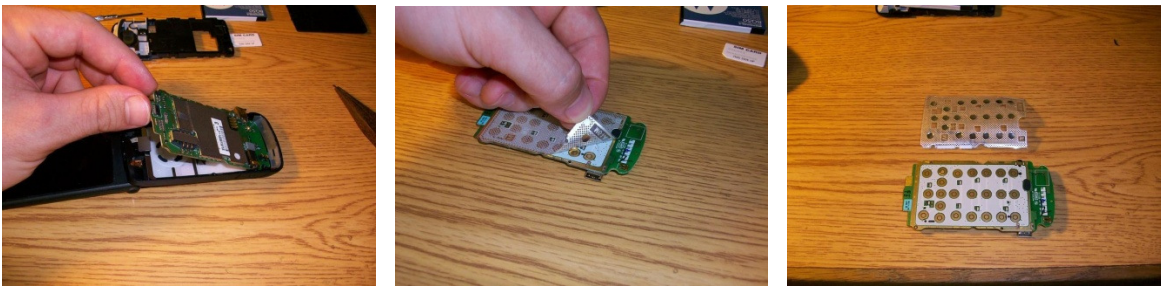
Fig 8: Isolating the W260g Motherboard

Pressing keys on the exposed circuitry of a cellular DTMF keypad is not as forthright a task as it might seem. The essential idea is simple: close a circuit across the leads of a given key. However, achieving this functionality physically and abstracting it such that a microcontroller can manipulate the keys is not so simple.

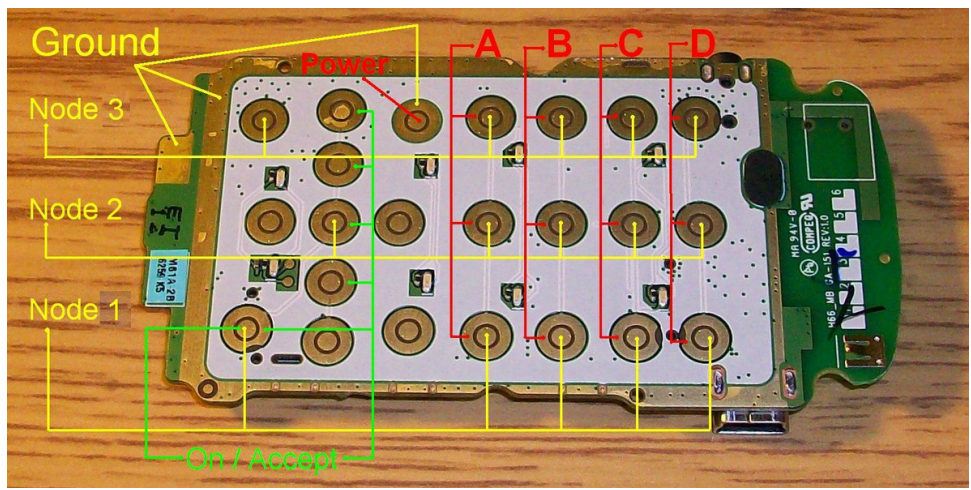
Fig 9: Electrical Nodes

The image above is a useful aid to understand the requirements for building the key-press circuitry. The electrical nodes on a cellular keypad are crossed in a grid, where the outer leads of the keys on the DTMF keypad are connected in 4 horizontal row nodes and the inner leads are connected in 3 vertical column nodes. Pressing any given key is a process of connecting one horizontal node with one vertical node. The convenient consequence of this is that any key on the DTMF keypad can be pressed by closing a circuit across a combination of a horizontal and vertical node. Thus, with only 7 connections, any key on the keypad can be pressed. The following chart indicates the combinations required to press any key on the DTMF keypad.

| Node | 1 | 2 | 3 |
|---|---|---|---|
| A | 1 | 2 | 3 |
| B | 4 | 5 | 6 |
| C | 7 | 8 | 9 |
| D | * | 0 | # |

Fig 10: Nodal Combinations and Keys Pressed

In general, cellular phones are not intended for the purposes required by this project, and so it is extremely easy to destroy the motherboard. The leads that form the keys on the keypad are made of copper, and thus solder is extremely adhesive to them. Unfortunately, the positive and negative leads for any given key are only millimeters apart, and accidentally soldering them together is the practical equivalent of destroying the board altogether. The key is forever frozen in a pressed state unless the solder can be removed. Since solder-wick is made of copper, it is rendered useless in this situation. A combination of careful soldering and some creativity with fork-terminals was used to build the appropriate circuitry to break out the nodes on the keypad for external manipulation.

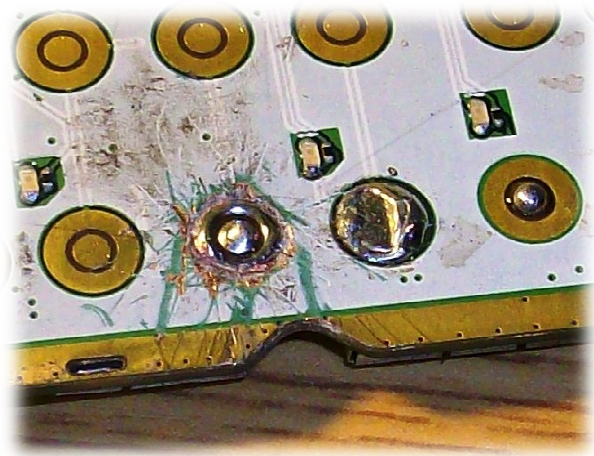

Fig 11: Aftermath of Trying to Undo a Mistake

The solution to breaking out the nodes of the keypad was arrived at after unfortunately destroying 4 different cell phone boards in similar fashion. The image above is the aftermath of an attempt to physically grind the solder off the nodes of the keypad—it's just not easy! Explicit directions to be wary of such potential mistakes must be included in any tutorial literature generated as a result of this project. The image below shows the soldering which was incorporated into the final design.

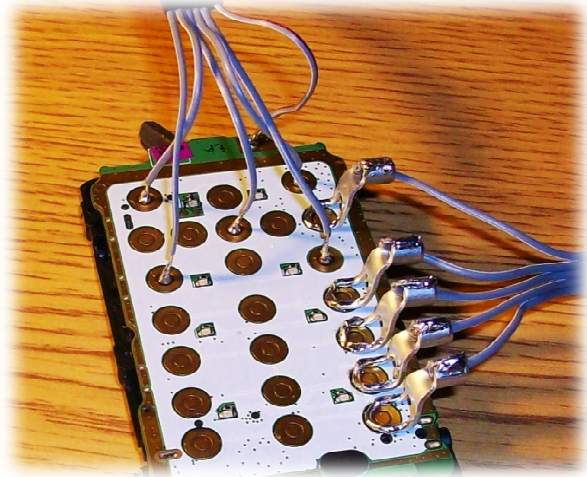

Fig 12: Properly Broken-Out Nodes

3.2.1.2.1 Why DTMF?

Whenever a call on a cellular phone is connected, the keypad is enabled to generate DTMF tones. DTMF— or TouchTone—is a well-established way to transmit information via audio. It provides 16 different tone values which can be employed to encode data, including of course the values of 0 through 9. Since the only parameters of concern in the transmission of remote monitoring data are numeric, this is all that is required. Furthermore, using an older standard like DTMF instead of binary data is actually a more realistic approach when considered in the context of what forms of wireless communication are actually available in Sub-Saharan Africa. While the continued proliferation of cellular technology in many places has made the expectation of cellular service in remote areas realistic, unfortunately, data networks are far less common. The reliable services to expect are SMS and DTMF.

3.2.1.2.2 Switching Take 1: Tri-State Buffers

Once the nodes on the cellular keypad are broken out, electronic switches must be installed to allow the microcontroller to manipulate them with control signals. The first approach used in this project was tri-state buffers. Tri-state logic basically allows for two wires to be connected or disconnected (set to high-impedance, really) by the assertion of a control signal. Unfortunately, when using tri-state buffers, the control signal is not electrically isolated from the load being switched. Furthermore, tri-state buffer chips have power and ground pins, which means that the parallel buffers on a given chip are not electrically isolated from each other. This created an odd phenomenon where every key on the keypad behaved as though it were drawn to ground whenever it was pressed, and as a result every key on the keypad registered only one of three possible values—1, 2, or 3. Since the mechanical action of pressing a key on a keypad represents an electrically isolated event, a proper electronic switch in this case must be isolated from the load.

3.2.1.2.3 Switching Take 2: Optoisolators

Optoisolators ended up being a ready and cheap solution to this problem, and the finalized circuitry in this project involves the use of 10 OPI110 optoisolators. The keys on the cellular keypad are center-positive, so the trick was to first connect the 3 column nodes to the positive pins on the load-bearing sides of 3 optoisolators. Then, in a fan-out style circuit, the negative side of each of these optoisolators was connected in series to the positive terminals of the rest of the optoisolators connected to the DTMF keypad. The 4 row nodes on the keypad are then connected to the negative sides of these optoisolators. Then, by asserting the control signals to switch one positive node and one negative node at a time, the logically connected key on the keypad is pressed. The same circuitry can be applied to switch the power and accept buttons.

3.2.1.2.4 Pressing Keys

Describing in words the way keys are pressed via the keypad circuitry can be confusing. A diagram is more useful in this case to understand precisely how this process works. The circuit diagram on the next page is a simplified subset of the optoisolator circuitry.

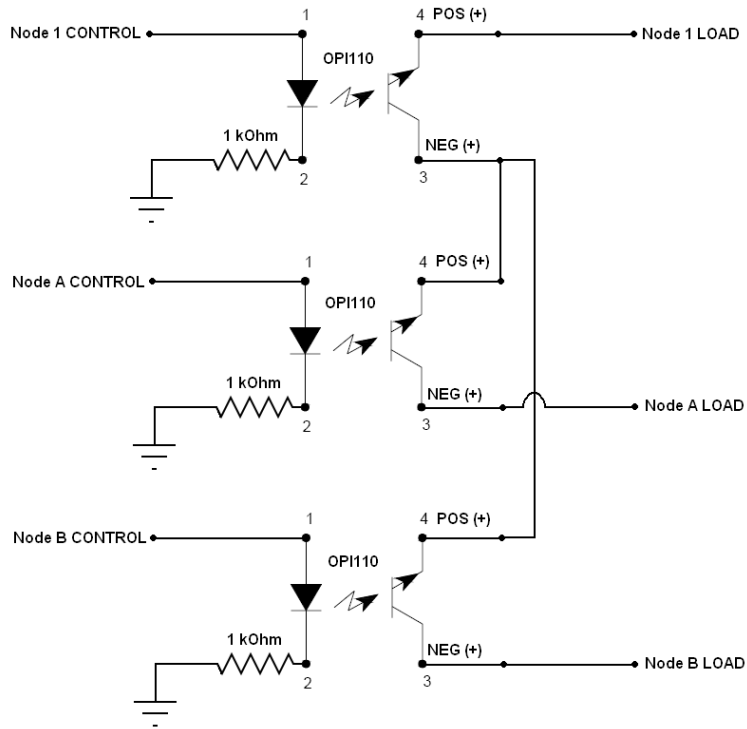

Fig 13: Optoisolator Circuitry for 1 Column and 2 Row Nodes

The critical thing to realize about the circuitry in this diagram is that when one numbered vertical node and one lettered horizontal node are asserted via the 'CONTROL' signals, then two nodes cross and a key-press occurs. In the above example, if 'Node 1 CONTROL' and 'Node A CONTROL' are both asserted high, then 'Node 1 LOAD' and 'Node A LOAD' will be *connected*. This would register as the number 1 on the keypad. If 'Node B CONTROL' were asserted, this would register as the number 4. This same circuitry is utilized for every button control on the keypad. Once this hardware is in place, the Freeduino I/O pins can be easily connected to the 'CONTROL' signals and by simply driving them high or low in the correct combinations, any key on the keypad can be pressed.

3.2.1.2.5 Incoming Call Circuitry

There is one tricky aspect to directly manipulating a cellular phone with a microcontroller, and that is the detection of an incoming call. When the monitoring software initiates a data transfer from a remote location, how does the Freeduino know? Whenever a call is received, the phone rings, and ringing occurs because a speaker circuit is connected to the motherboard of the cell phone. The trick to detecting an incoming call is the ability to drive an input pin on the Freeduino high whenever a call occurs. This can be accomplished using the same circuitry implemented for key-presses, except in this case the control signal on the optoisolator is connected across the positive and negative terminals of the speaker leads. This circuit is shown below:

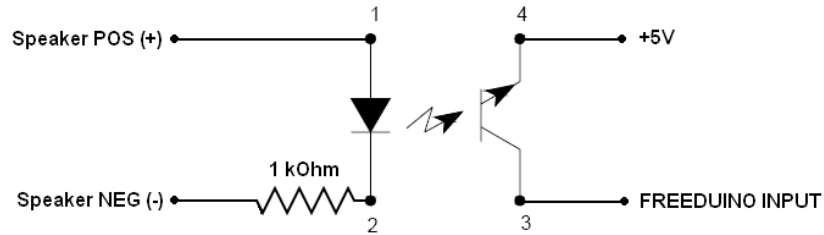

Fig 14: Incoming Call Circuit

When an incoming call is received, current flows between the positive and negative terminals of the speaker, lighting the diode in the optoisolator. The +5V source from the Freeduino board is then switched to an input pin on the Freeduino, driving it high. The Freeduino thus has an input which can fire an interrupt, and it can respond in turn by pressing the 'Start' button, accepting the call. This interrupt is discussed in a later section.

(Note: Ironically, this is the same circuitry infamously used by terrorists to remotely detonate bombs with cell phones. All that would need to be modified here is that instead of the Freeduino input there would be an electrically triggered fuse. Of course, this just goes to show that technology is the same whether it is used for good or evil.)

3.2.1.2.6 Outgoing Call Circuitry

When an incoming call occurs, an interrupt is fired by asserting an input on the Freeduino, and the response is to connect the circuit that controls the 'Start' button. Quite uninterestingly, the process of dialing out on the phone involves simply dialing the number by manipulating the keypad and then re-asserting the 'Start' button. And, of course, the termination of any call is accomplished by asserting the 'Power' button.
The entire circuit that is built onto the motherboard of the cellular phone has now been explained. One important thing to note at this point is the generic nature of this hardware configuration. Most cellular keypads are implemented nearly identically to the board described in this section. Furthermore, all phones are equipped with a DTMF keypad, a start button to accept calls, and a power button to terminate calls and toggle the phone on and off. Thus, the implementation of the hardware that interfaces with the Motorola W260g here is highly reusable. This circuit will work with *any* keypad that has a sufficiently similar scheme of asserting key-presses through a mechanical connection of two conducting leads.

One item to note, however, is the fact that this phone is conveniently charged via a 5V USB cable, which enables the entire board to be powered without the battery by simply soldering a wire from the positive battery terminal on the cellular motherboard to the +5V power output on the Freeduino. If a future cellular phone required the more modern USB standard of 3.3V, this could also be accommodated by the Freeduino. This may not be generic to most phones, unfortunately. It is important before attempting this sort of power source re-engineering to accurately establish the safe operational voltage limits of a given cellular phone.

*3.2.1.3 The Freeduino Main Control Logic*

Once all of the hardware has been constructed and the Freeduino is able to press keys on the cellular keypad by asserting output pins, the next part of this project is the implementation of the logic to make the solar controller and the cellular phone operate in conjunction with the microcontroller.

The Arduino compiler requires that two functions be implemented for any program to be uploaded to a board. These are fairly intuitive and straightforward: there is a `setup` method and a `loop` method. The intent of these methods is equally intuitive. `setup` is used to initialize I/O pins, serial connections, and any other necessary state, and `loop` is the main control loop which is executed for the duration of time the board has power. `setup` is called once and only once at the beginning of the program.

The function of `setup` in this case is straightforward with one exception. After all other initializations have been completed, the `setup` function then asserts the power button on the cell phone for a period of 4 seconds, and then waits for 20 seconds. This action is required to turn the phone on and allow it to reach a state where it is ready to receive calls.

3.2.1.3.1 Main Control Loop

Once inside the main control loop, the logic of the program is fairly straightforward.

```
void loop()
{
  // check if time to do a transfer
  if( incoming call interrupt has fired )
  {
    // output our data frame
    start the call;
    output current data;
    terminate the call;
  }

  // poll for data every 30 seconds
  if( 30 seconds has not elapsed )
  {
```

```
      wait briefly;
  }
  else if( 30 seconds has elapsed )
  {
    // update the data fields
    log and queue old data;
    poll the controller for new data;
    validate polled data;
  }
}
```
Fig 15: Main Control Loop Pseudo-code

The above pseudo code is of course abridged, but the general idea is intact. An externally fired asynchronous interrupt will set a flag that is checked in the loop, and when it is set we know that a call from a remote server has been initiated. The call is connected, and a queue of data is output to the cell phone.

The rest of the loop is concerned with updating the current data values. Once every 30 seconds (an arbitrary number), the microcontroller will poll the solar controller for updated information on about the solar power system. The specific data parameters are discussed in more detail later. This process is not implemented as a timed interrupt because it is important that these two processes happen synchronously, i.e. if a call is initiated and data is being transferred, the update function cannot interrupt the control flow, and vice versa an incoming call cannot interrupt the process of updating the current system data. Once one of these processes has begun, it will proceed until finished.

The logic for the updating the current system data is the only slightly interesting part of this code snippet. When the solar controller is polled, the data received is validated. If the voltage of the battery bank is found to be lower than a pre-set cutoff threshold, it is assumed that the bank is being overdrawn and an alarm is started. This alarm occurs in the form of a call to a specific number, which should ideally be the number of the system's primary overseer. This alarm has the potential to be thrown on every update, so if there is a problem, the system will sound an alarm continuously until appropriate action is taken.

The other interesting part of this code is the logging and queuing functionality. This is intended as a cost-saving mechanism. Every time a call is initiated, a minute of paid time is used up by the cell phone (even if the call is less than a minute). Thus it behooves us to queue updates and retrieve them in batches that can take up to but never exceed one minute of time to transfer. Since the maximum number of data frames that can be guaranteed to fit in the span of a minute has actually been determined to be no more than three, the queue is made of only two previously stored data frames, plus the most recently polled data. This may seem slow, but this is because of the behavior of the DTMF keypad—regardless of how quickly a set of keys are pressed, the tones will be buffered and sent over the line at a constant rate of roughly two tones per second.

Since every call will return three full frames of data, if an implementation were set to queue up a log of system state every hour, we would only need to call the device once every three hours to receive the equivalent of hourly historical data. Historical data is not time-sensitive in the same way that a low-voltage disconnect alarm requires immediate action (turning off the load or the inverter). In the case of historical data, it may take some hours to see the values stored, but in the case of an emergency, system overseers will be notified within 30 seconds. This is the "no news is good news" approach. Performance is a secondary concern to price in the case of remote solar monitoring, and that is why the obviously slow nature of this system specification is acceptable.

3.2.1.3.2 Transfer Protocol

The idea of data frames has been mentioned already in the description of this project. When the cellular phone is called and data is being transferred as DTMF tones, it is sent in organized messages called frames. The decision of what to put into these messages was a long and thoughtful process, and many interesting ideas were entertained along the way.

The original idea was to utilize all 16 DTMF tones and encode all data being sent over the call in a binary representation. With 16 different tones, each one can represent 4 bits of data, i.e. 0 = 0000, 1 = 0001, 2 = 0010, and so on. This system would allow any binary string to be encoded as DTMF tones, where each byte is the conjunction of two tones in sequence. The software that receives a transmission of audio information in such a format would simply decode the audio tones and map them to the appropriate binary values.

There are problems with this system, however. The DTMF tones A, B, C, and D are not part of a standard keypad and are often reserved for special use. An unfortunate combination of these tones could cause a call to be disconnected or connected to an emergency phone line used by, for instance, the police. There is no way to prepare for such an instance because these protocols are, for obvious reasons, hidden from the public.

With only 12 keys available on the key-pad of the cell phone, unless DTMF sinusoids were being physically synthesized by the microprocessor, the next available options to encode data as binary include the use of an octal representation (8 keys representing 3 bits each) or a 4-key system in which each key represents 2 bits of data. With the discovery that the cell phone buffers tones that are pressed into it and outputs them over the call at a constant interval, every possible way to encode data as binary was basically thrown out. It simply required too many tones. One of the cost saving mechanisms that this project must consider is how to get the most information across in a call in the least amount of time. As it turns out, simply using all of the DTMF keypad and transmitting the polled data verbatim as decimal values is the optimal solution.

But even doing this requires a protocol. What is the order that tones should be transmitted? How are they distinguished? What are the units? There is no period on a DTMF keypad, so decimal point values need to be accounted for in such a way that can be easily translated back and forth. In the end, simplicity is preferred, and it was decided that the best way to build a data frame was to just use the '#' key as a field separator, and then designate that for any given field the right-most two digits are *behind* the decimal point. Ergo, the string '#01#' does not mean 1, it means 0.01. (During a system test, this value actually showed up for the panel current at one point as a result of moonlight.) In the end, the order and fields in a given data frame were decided upon as follows:

```
                # PANEL VOLTAGE (V)
                # PANEL CURRENT (A)
              # BATTERY VOLTAGE (V)
           # BATTERY CHARGE CURRENT (A)
          # TOTAL KILOWATT-HOURS (kWH) #
```

Fig 16: Project Data Frame

3.2.1.3.3 Interrupts

The way in which the Freeduino takes the appropriate action when receiving an incoming call signal is through the use of an interrupt. The Arduino environment actually allows for the easy integration of external interrupts on pins 2 and 3. In this case, when the interrupt for an incoming call is fired, because of the nature of the oscillating voltage on a speaker, the interrupt is actually thrown many times, and some means of mutual exclusivity is actually important to prevent the interrupt flag from being thrown again after a call has been accepted and the flag has been cleared. The code that completes this task is shown below:

```
void incomingCallISR()
{
  static int mutex = 1;
  if( mutex == 1 )
  {
    noInterrupts();              // disable interrupts for critical section
    --mutex;                     // toggle the mutex -- stops other execution
    setExecuteTransfer( true );  // enable data transfer
    ++mutex;                     // toggle the mutex -- releases control
    interrupts();
  }
}
```

Fig 17: The Incoming Call Interrupt Service Routine

3.2.1.3.4 Alerts and Validation

An update in this program occurs every 30 seconds. When the controller is polled for the most recent values of the diagnostic parameters of the solar power system, these values must be validated to determine whether or not a problem has occurred. In this case, the low voltage cutoff for the battery is tested for, and if the voltage is too low, an alert is initiated in the form of a call to a specific person.

```
if( currBatteryVoltage <= LOW_VOLTAGE_CUTOFF )
{
    // an error has been detected. alert the masses!
    alert();
}
```
Fig 18: Error Detection

*3.2.1.4 Implementing RS-232 Communication*

The implementation of the interface between the Freeduino and the TriStar-45 was an interesting and challenging problem. This process also revealed some critical weaknesses of the approach taken by this project to avoid dependencies on specific hardware and communication protocols.

3.2.1.4.1 Implementing Modbus RTU

The implementation of the controller transfer protocol is a different software module than the main control software described above. An Arduino software library called `ControllerTransferProtocol` was created to enforce this separation. This library allows an instance object to be queried for several items of information, and includes such functions as `getBatteryVoltage`, `getPwrSrcVoltage`, `getChargeCurrent`, `getLoadCurrent`, and `getTotalKilowattHrs`.

The way that any one of these functions works is the same. The Morningstar TriStar-45 controller can be polled using the Modbus protocol. There are two flavors of Modbus, the ASCII specification and the RTU (Remote Terminal Unit) specification. The TriStar-45 uses the RTU specification of Modbus.

The difference between the ASCII and RTU specifications of the Modbus protocol is that the ASCII specification separates fields with the colon character and the RTU specification is based on timing. At the baud rate of a serial transfer, the RTU specification requires 4 byte-lengths of idle time before receiving a new command. Then, the data fields of a command must be separated by less than 2 byte-lengths of time. This is fairly convenient; the protocol ends up manifesting itself as simply waiting the appropriate period of time, and then transferring all of the bytes of a message over the line as fast as possible.

There are several functions in the Modbus protocol, but the only one of them that this project is concerned with is called the "Read Holding Registers" function, and it is named using the binary equivalent of hexadecimal 0x03. The TriStar-45 has around 20 parameters which are continuously updated and held in data registers. Among these, of course, are the items of information alluded to in the operation names above like the panel and battery voltage. In order to call the Read Holding Registers function, we need the address of the device we want to poll, (when there is only one device to communicate with, the address of the TriStar-45 is 0x01), the code of the command (0x03), the address of the starting register we want to read, the number of registers to output consecutively after this register, and then a 16-bit Cyclic Redundancy Check which is used to test that the message was received intact. In order to read, for instance, the voltage of the battery bank, we transfer the following string of bytes to the TriStar-45:

```
0x01 0x03 0x00 0x08 0x00 0x01 0xC8 0x05
```

Addresses are 16-bits in the TriStar-45, and this is why the bytes 0x00, 0x08 (0x0008) and 0x00, 0x01 (0x0001) are used, respectively, to address the battery voltage holding register, and to indicate that we only wish to receive a single register's worth of data in a response frame. The CRC generator uses a polynomial function based on the bytes of the message to generate the last two bytes of data, 0xC8 and 0x05. This is a complicated process that can be better understood by reading the documentation, and is not particularly relevant to discuss here.

Response messages are equally simple. They are composed of the device address they are from, the function code they are a response to, the number of bytes in the message, the data bytes in order, and a 16-bit CRC value. The response to the previous function call, if the voltage of the batteries were 12.53V, is the following:

```
0x01 0x03 0x02 0x10 0x98 0xB4 0x2E
```

The only value of real interest here is 0x1098, which is the combination of the 4[th] and 5[th] bytes of the message. This is a binary integer, which means nothing until it is multiplied by a scalar, which is specified in the Morningstar documentation. The value that this integer must be multiplied by is 0.0029500042724609375. The result of this multiplication is the double value 12.53, which is the voltage of the battery bank.

Every call to the controller is implemented in the same way, although the scalars for each value vary. Once the final double values have been calculated by the `ControllerTransferProtocol` object, they are returned to the caller. So, the result of `getBatteryVoltage` is simply a double, which in the case of the above example is 12.53. In the case of a different controller with a different protocol, the result would be the same to the calling logic, so the implementation of the transfer protocol is hidden from the main logic of the program that translates this value into a form that can be transmitted over the air waves as a string of DTMF tones. This promotes the decoupling of the details of the transfer protocol from the logic that uses and transmits the data polled form the controller to the remote caller.

3.2.1.4.1 Modifying the Freeduino Serial Port

The final and probably most unexpected hurdle in this project was the building of a new, customized serial port to allow for the proper interaction between the Freeduino board and the TriStar-45. The problem with simply plugging the two devices into each other became readily apparent early on because both the solar controller and the Freeduino have female-ended DB-9 connections, indicating that they are both DCE (Data Communication Equipment) devices. The usual consequence of this fact is that a null-modem connector is needed to cross over the TX and RX pins such that the TriStar-45 RX pin is receiving the output of the Freeduino TX pin and vice versa. But this turned out to not be sufficient. In addition to this fact, the Freeduino makes use of pin 4 on the serial port as an automatic reset pin, so whenever the two were connected together, this unfortunately resulted in the Freeduino board being reset continuously in a loop. The reason for this is the fact that the TriStar-45 serial port expects a positive RS-232 reference voltage as an incoming value on pin 4. This is because the circuitry that receives incoming commands on the TriStar-45 is optically isolated from the rest of the controller, and so it must derive the high voltage necessary to transmit and receive data from the polling device—this is known as "port powering". The diagram below shows the circuitry described in the Morningstar documentation for the TriStar-45 RS-232 port.

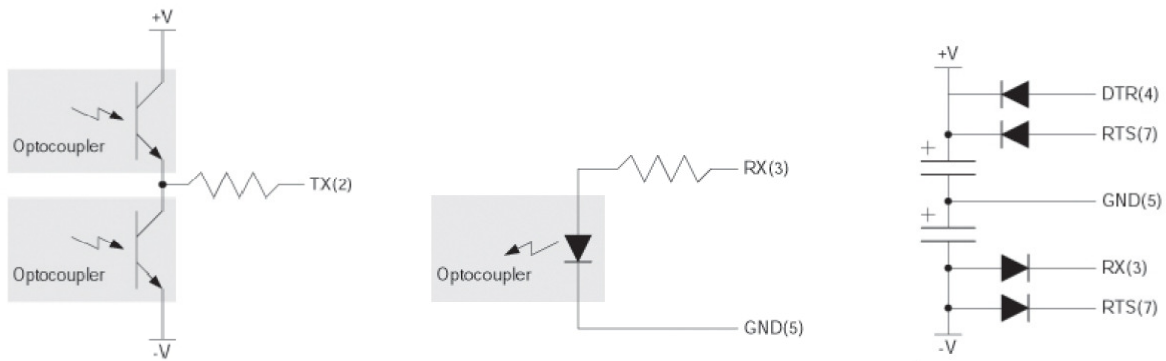

Fig 19: TriStar-45 Simplified RS-232 Circuit (Source: Morningstar Corporation)

Fortunately, some null modem cables cross pin 6 onto pin 4 in addition to crossing RX and TX, and pin 6 on the Freeduino is not being used at all. The diagram below shows the wiring that Morningstar Solar indicated in response to an inquiry should be used to power the TriStar-45 port from the Freeduino port.

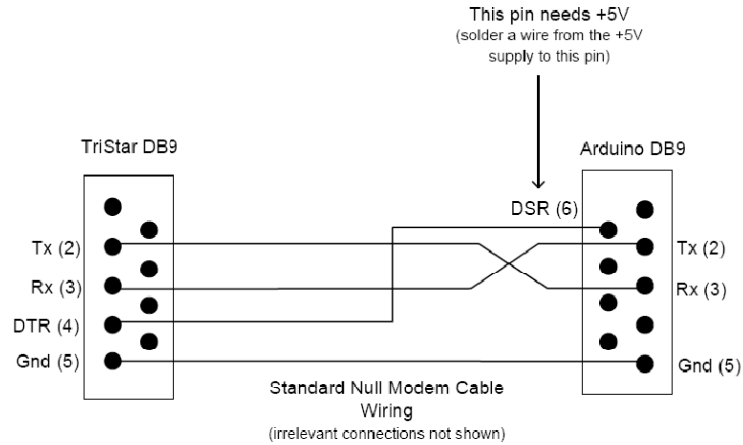

Fig 20: Port-Powering the TriStar-45 with a Null Modem (Source: Morningstar Corporation)

However, even this diagram has a problem, because it is generally incorrect to wire a +5V jumper into an RS-232 port, which has a positive transfer voltage of +12V. So, the quick fix to this problem was to route the VS+ pin on the Max232 chip of the Freeduino board (used to amplify the signal of the Freeduino to RS-232 level) to pin 6, and completely disconnect pin 4. Once these changes were instituted, the devices communicated as expected.

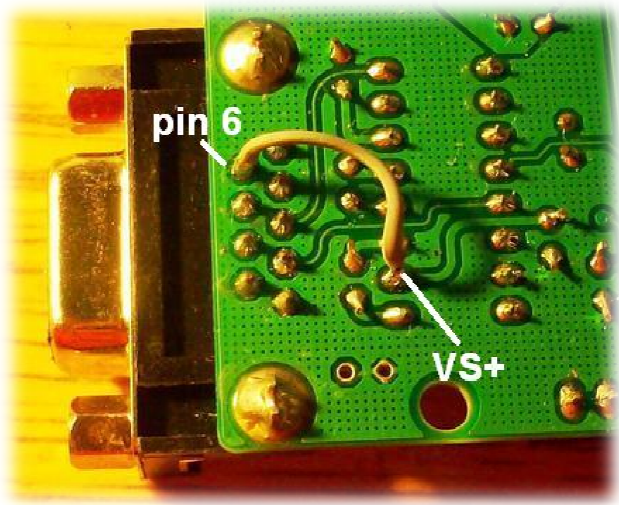

Fig 21: Jumping VS+ to Pin 6 of the RS-232 Port

There are some obvious disadvantages to this approach. First of all, there is an unfortunate amount of hardware coupling going on here. Despite the fact that the software implementation of the Modbus protocol is entirely separate from the main control software, the fact that a wire from the VS+ output of the Max232 chip had to be soldered into the unused DSR (Data Signal Ready) pin on the serial port of the Freeduino unfortunately ties this implementation specifically to the TriStar-45. If another controller did not use a port-powered circuit or made handshaking use of the DSR pin, this hardware specification would be incompatible.

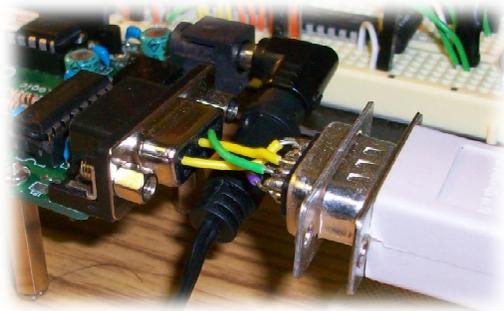

Fig 22: Customized RS-232 Port

The entire specification for the hardware component of this project and its functionality has now been described. The image below shows the complete implementation, with critical components labeled.

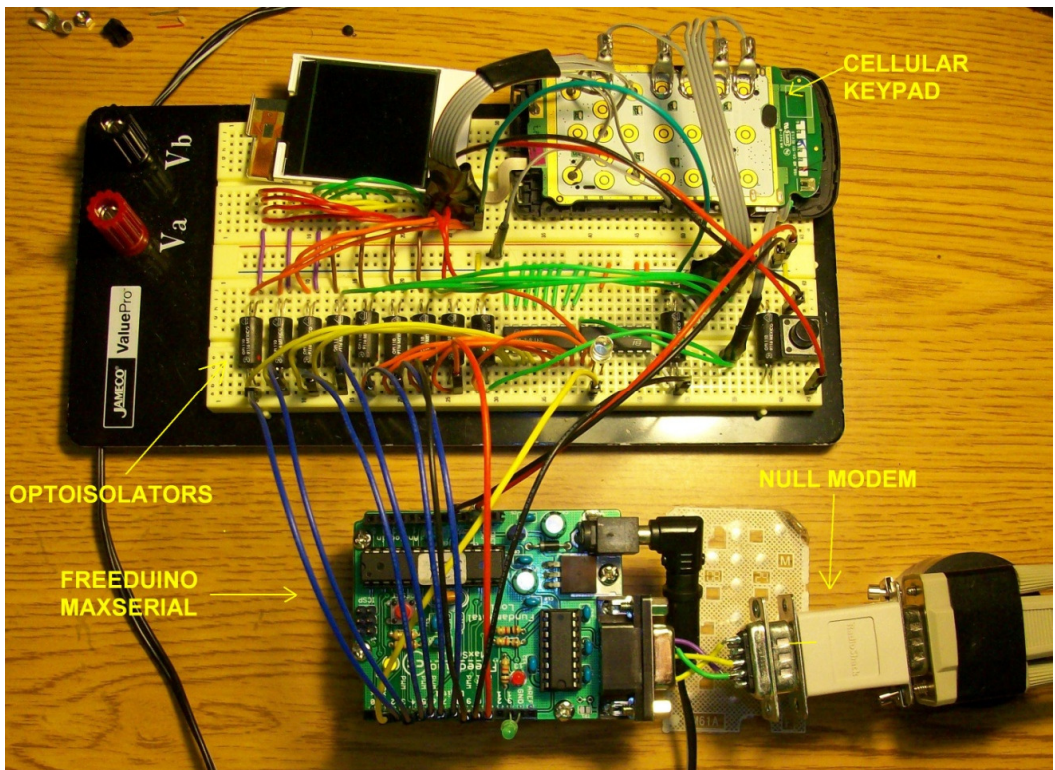

Fig 23: Completed Hardware Implementation

*3.2.2 Building the Software Component*

The requirements for the software system that should utilize the hardware device in this project have already been described. Unfortunately, time ran out this semester and no concrete implementations could actually be completed. Thus, there is nothing in particular to describe here, and the Future Work section of the conclusion is utilized to cover what possibilities exist for potential future implementations.

*3.2.3 System Test and Results*

It is unfortunate that the software side of this project could not be successfully implemented; however, the hardware component described above has been extensively tested. During the week of November 24-29, this project was installed on a working 250W solar power system, and every implemented functionality was found to work as expected. Preliminary testing with the TriStar-45 was done using a small deep-cycle battery, because the TriStar can function without being connected to a solar power array provided it has a constant voltage source. However, this has the obvious pitfall that only one value returned by the implementation of the transfer protocol could ever be tested. Once the system was integrated with a working solar power system, the value of the panel voltage during the day and the current output could be seen.

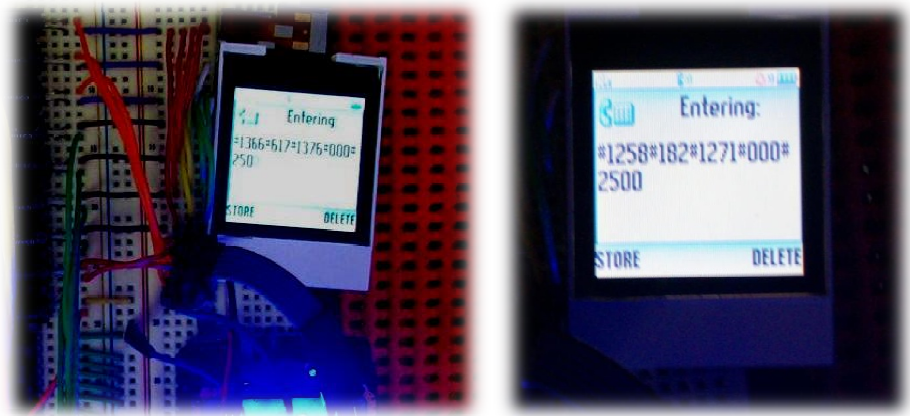
Fig 24: Outputs during System Test

These images, taken at separate times of the day, show the output of the cell phone monitor after a call has been initiated. The first image, taken at 10:00 AM, indicates that the voltage of the panels is 13.66V, the current is 6.17A, the voltage of the battery bank is 13.76V, and the battery load current is 0.00A. The reason this value is zero is the fact that the TriStar-45 has a 'Load Control' mode that allows it to control DC loads drawn from the TriStar directly, i.e. not through an inverter connected to the battery bank. Since this system does not use and DC loads, this value is always going to be zero. This field was left in the implementation because some controllers display the current that is being routed from the charge controller directly into the battery bank. The last value, which was output completely at the instant this image was taken, corresponds to the temperature of the TriStar-45 heat-sink, which defaults to the value of 25.00 C. This is a field that has since been removed from the data frame because it requires an external sensor on the TriStar-45. (This field has of course since been replaced by the total kilowatt hours produced by the solar power system.) The second image, taken at 11:00 AM the previous day shows the difference in system performance under poor weather conditions (it was overcast). The images below show the setup of the system test.

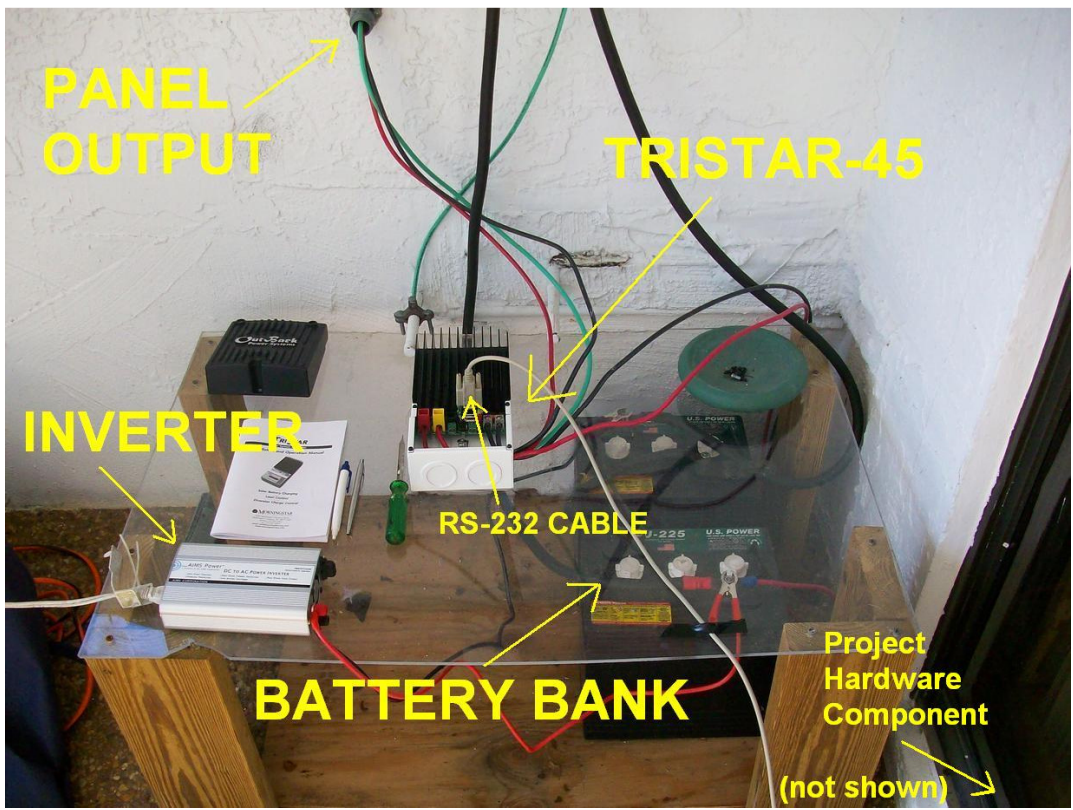

Fig 25: System Test Setup

The complete results of this test are recorded in Appendix A, but one diagram corresponding to the results of an extensive test on 11/27 deserves to be shown here. The graph below was manually created from data recorded during the test.

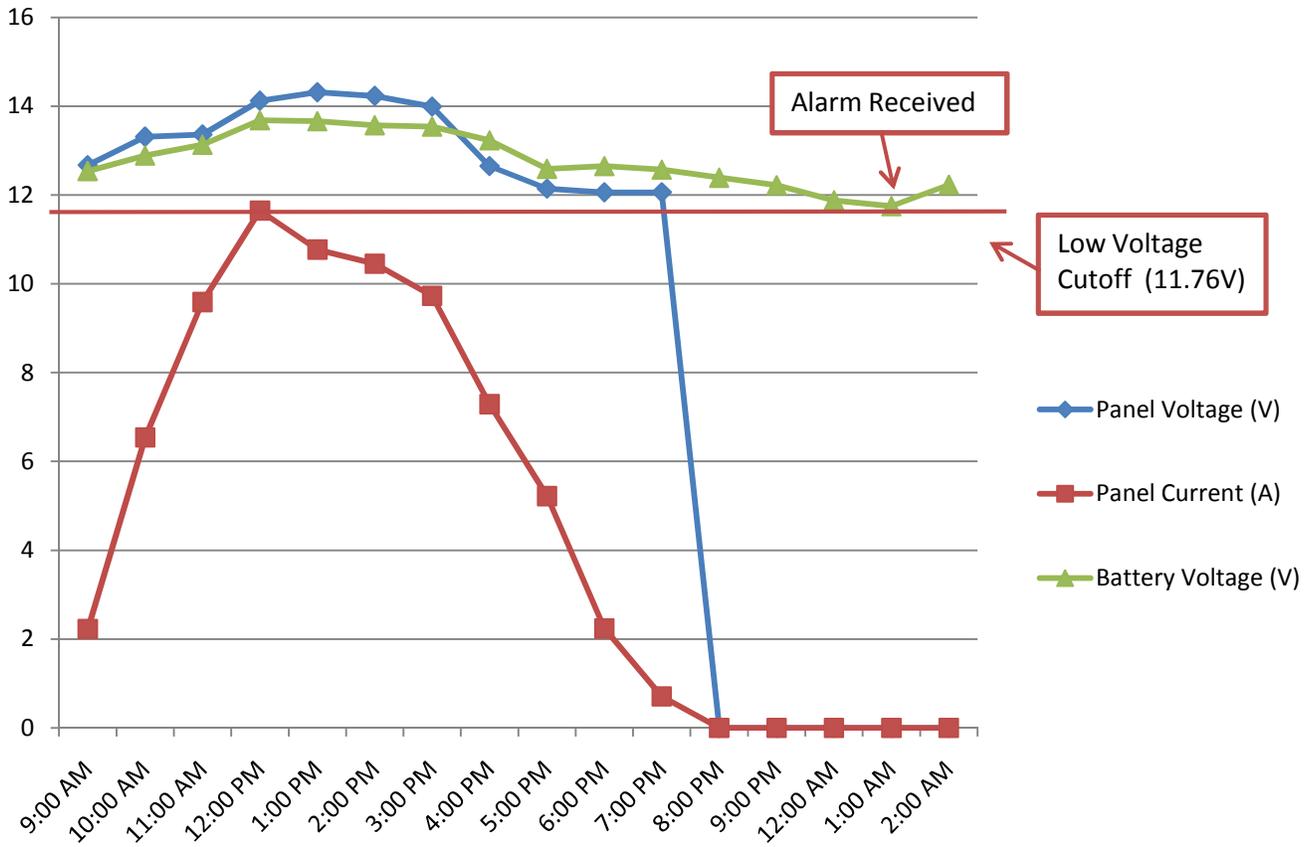

Fig 26: 11/24 System Test Results

This test demonstrates the full functionality of this project. During the day, on the hour, from 9:00 AM to 9:00 PM (less often afterwards), the cell phone was called, and the results were recorded. At night, the inverter was switched on and a 300W load was run from 9:00 PM until around 1:00 AM, when a call was received because of a low voltage cutoff detection. This demonstrates that the fundamental alarm functionality of the system works as expected, and the graph created here is an example of the potential for logging and analysis software to utilize the output of this hardware system.

This concludes the discussion on the implementation of this project. In the Conclusion section of this thesis, the lessons of this project and the potential this has to be expanded into a framework for further applications are discussed.

# 4. Conclusion

This project has some significant claims to success as well as some admitted failures. The greatest success was the completion of the hardware component of this project. The most obvious failure is the fact that the server-side software that polls and stores data obtained from the monitoring device was not completed. However, it can be argued that while this may have been a desirable accomplishment, in essence the software side of this project is a fairly straightforward VoIP-telephony web application equipped with some DTMF decoding software. None of the requirements of the software side of this project represent new or particularly interesting problems. All that is really required of a software application that uses this system is the ability to decode DTMF tones. In some ways this simplicity could be considered a strength of the approach taken here. Nevertheless, the accomplishment of the hardware side of this project represents the critical proof of concept for this system because not only does the hardware represent the most significant initial cost in any remote monitoring system, it also represents the source of most after-installation costs. Thus, the ability to show that this implementation accomplishes the requirements of this project is much more contingent on the success of the hardware component than the software component.

## 4.1 Why This Implementation Is Viable

The advantage of this system is above all its simplicity. Upon testing this system with a real solar power system, it became clear that in some cases the essential function of alerting system overseers when the battery bank is drawn too low may actually be all that is required or practical. After all, this can be accomplished at almost no cost at all after installation. Whenever a problem arises, an alert fires, and assistance is called for. A solar system can function without the logging aspects of monitoring if necessary, but it cannot operate if it is being progressively compromised by unintended misuse. In the context of Sub-Saharan Africa, in some situations, this may be the only affordable setup.

This project has the added advantage of being inexpensive. Using a cellular phone as a transmitter is obviously not a conventional approach; however, the effort here was to demonstrate that the theoretical minimum cost to accomplish the task of remote monitoring is in fact orders of magnitude cheaper than existing systems would indicate. The cellular modems on the market that can interface with the TriStar-45 out of the box are on the order of $200-$800 dollars, and monthly service fees for monitoring services are on the order of $50 a month. The Freeduino board, when purchased unassembled, is $15, and the Motorola W260g is available for an additional $14-$15. Combined with the optoisolators, wires, and other circuit elements used here, this entire system can be built for no more than $60 dollars. This is half of the cost of a single GSM

chip, and a fraction of the cost of the cellular modems utilized by existing solar monitoring systems. When considered in the context of Sub-Saharan Africa, the solution proposed by this project is a vastly more affordable option for remote monitoring. A more robust and potentially expensive system could be developed using this project specification, however, the bottom line here is that up-front costs have been reduced to a fraction of their current market value, and after-installation costs have been nearly eliminated. Not only does this project meet the information needs of effective solar monitoring and maintenance, it has seriously addressed the problem of cost. This is obviously going to be too expensive in certain situations. It is doubtful that remote monitoring can be accomplished for less, however, but this is a limitation of remote monitoring in the first place. This project is open to such criticisms.

Finally, this design is resilient. I have field tested this device in conjunction with a real solar power system for several days on end and it exhibits no debilitating flaws or obvious points of failure. If problems occur it accurately and reliably detects them and raises the appropriate alarm. When things are working properly it accurately reports the parameters it polls in DTMF sequences. When errors do occur in the form of a missing tone here or there, it is usually because of a lack of cellular signal quality.

## 4.2 Why This Implementation Is NOT Viable

There are some obvious disadvantages to this approach. First of all, as has already been discussed, there is an unfortunate amount of hardware coupling going on between the Freeduino and the TriStar-45. A better argument for the viability of this approach would include the demonstration that this design can work with multiple controllers and cell phones.

The next obvious point of contention with this design is the practicality of finding the parts to build an Arduino board in Sub-Saharan Africa. The availability of the specific parts utilized by Arduino boards in different parts of the world is admittedly unknown. One part in particular that may not be so easily found is the actual microprocessor, the ATMega328. While AVR microcontrollers are easily obtained and cheap in the United States, this may not be the case in, for instance, Zambia or Namibia.

The third issue which must be critically examined in any future development effort is the sustainability of this system's power supply. It currently runs off a battery, which will eventually die. Some circuitry must be introduced to enable this battery to be routinely recharged without becoming a phantom load on the battery bank, as this is a huge issue in the depletion of deep-cycle batteries. Small, continuous loads can wind up destroying a battery bank long before it would otherwise functioning.

Fourth, and this is not a theoretical disadvantage of the concept demonstrated by this project but rather a consequence of the specific implementation, if it were of greater practicality to use the Arduino board to synthesize DTMF sinusoids, and simply transfer the tones audibly over the line, the only applicable units of code here are the ones implemented without respect to any output pins. The TriStar-45 Modbus RTU implementation will still work because that is agnostic with respect to the pin configuration. It deserves to be noted that DTMF synthesis is an unnecessary and error-prone exercise, especially considering the fact that there are almost no cellular phones in existence without a 12-key DTMF keypad, an accept/start button, and a power button. This implementation is the most extensible. A different circuitry may be required for some oddity of a specific cellular phone, but the abstraction of a nodal grid and the use of optoisolators to separate load switching from control signals is very modular.

Fifth, while the C++ class developed to implement the TriStar-45 Modbus protocol is a good example of encapsulation and implementation hiding, the main control logic of this program is implemented on a very low level because of the embedded environment in which it runs. However, the code for this entire system uses only about a $5^{th}$ of the EEPROM in the Arduino to store and run this entire program. The Arduino environment actually provides some capability for object orientation, and breaking up this design into more Arduino C++ libraries is probably one way to allow further development to localize complexity and implement only the required logic for a given controller or external circuitry. This will increase the memory footprint of the program, but there is plenty of room to grow at this point.

Sixth, it might be practical to use a more robust hardware implementation or different wireless communication medium in some cases. If it is affordable or more practical, sometimes the use of a more powerful microprocessor can be of some advantage. Furthermore, the use of SMS messages may be more practical as a means of communication with individuals and software systems. Many programmable GSM processors have the ability to send SMS messages, and this is advantageous over audio DTMF for two reasons: 1.) It is cheaper to send SMS messages than to actually call, and 2.) sending SMS messages is a lower-power process than calling, and may be preferable if the power consumption of the current implementation is unacceptably high. Regions with data networks also offer the opportunity to use more robust communication, and, if this is a practical alternative, then this system is a bit primitive. It should be noted, of course, that the current level of simplicity in this project and use of such anachronistic paradigms as DTMF is viewed as a preferable way to go because of its still-pervasive use in cellular telephony networks.

The last point of contention in the acceptance of the results of this project concerns the fact that this has *never* been tested in the field, in any form. Remote monitoring has never entered into the design considerations of any solar energy dissemination programs in Sub-Saharan Africa, and the practicality of introducing it may be one particular reason for this. Building this system, even if it is open source, requires a significant background knowledge of electronics, microprocessors and programming. As it is routinely cited that lack of training in the maintenance of solar power systems is one of the main causes of failure, a development initiative involving a system of this complexity may be cumbersome and problematic. Technology is seldom if ever a solution to the systemic problems of underdevelopment, and it is likely wishful thinking to expect dramatic changes as a result of the integration of the results of this project into a future solar dissemination program. However, it may help in a few cases.

4.3 Future Work

The obvious thing to consider in the future development of this project is how to design an open-source software system to poll and log the data obtained from the remote monitoring device detailed in this project. This was attempted prior to the writing of this thesis, with limited success. The problem is actually that the decoding of DTMF is *too* common of a telephony application, and it is almost never used in this fashion. Java's Telephony API (JTAPI), for instance, is specifically geared to be used on top of a robust PBX system in an industry-scale call center implementation. Furthermore, while such open-source telephony frameworks as Asterix are incredibly complicated and powerful, it is completely overkill for this application.

One implementation that showed great promise was the use of Skype's Java API. While this is a paid service when calling land-lines or cell phones, Skype is a fairly straightforward and common VoIP telephony application, and a lot of open-source development is actually taking place using its APIs. The problem with this, however, is that DTMF detection is not well supported. It is likely that DTMF event detection will just have to be implemented from scratch to avoid the massive overhead of using call-center APIs.

Other future work for this project has been outlined indirectly in the previous section. Limitations on this design are to be considered critically, and further development of this project should take into account changes that might make a given implementation more suitable to the environment in which it is intended to operate. One of the most promising future changes as far as the hardware side of this project goes would be the integration of SMS messaging. There are multiple ways to accomplish this. Either a GSM chip could be equipped with the application logic to send intelligent SMS messages, or, more simplistically, the button presses required to send any message (i.e. an 'n' is two consecutive presses of the '6' key provided the phone is initialized into the correct state) could be built into the software and data could be transmitted in the same way a human would attempt to use a phone to send a text message. This could likely be accomplished by the microcontroller faster because key presses are timed to be slow in the case of pressing keys to generate

DTMF tones. There are of course obvious problems with this. Not every phone implements texting the same way, some phones have extensive menu lists for non-alphanumeric characters, some phones implement the space character using the '0' key instead of the '#' key, etc. This makes the abstraction of text messaging difficult to develop in such a way that does not have to be changed for every phone. Most every phone, again, has a DTMF keypad, a power button, and a start/accept key. Thus, the best way to implement an SMS system that can be transplanted and reused is likely to involve the use of a programmed GSM chip. This, however, carries with it the great pitfalls of using a proprietary piece of hardware and tying an application to the company that manufactures it.

4.4 Review

The general rule, going forward, must be that if this application is to be developed and expanded upon, implementations must be carefully abstracted so as to make as few implementation-specific design choices as possible. Allowing individuals to customize applications without rewriting this project from scratch is a powerful concept, but of course this must walk a fine line between useful design assumptions and the potential confusion that may result from leaving too many critical details up to an implementation. As with any framework, sometimes making a project easily extensible for the majority of potential users is preferable to trying to accommodate every imaginable scenario.

It is with the latter context in mind that this thesis was titled as an open source 'framework' while throughout the course of this paper it has simply been referred to as a 'system'. This project is not the solution to the larger problems described in this thesis—it is a proof of concept. The expansion of this idea into other applications using other devices, and further development in general is the reason that this should be thought of as the beginnings of a framework rather than a working, ready-to-go implementation. The fundamental premise established by this paper is that solar power systems are implemented as a way to promote development in Sub-Saharan Africa, and they routinely fail because of poor maintenance. Remote monitoring is a way to assist in maintenance process, but it is too expensive. This project is the fundamental proof that it does not *have* to be, and can in fact be done effectively at considerably low cost. As there are many applications and potentially different implementations out there, the most important contribution that this project can claim is the initiation of a hopefully collaborative and robust future development effort.

# Acknowledgements

I apologize in advance to anyone I may forget in this list, as it represents a history of close interactions with so many different people that I may, in my haste, neglect to mention a few. While I would prefer not to overly conflate the intent of this project with its timeliness and significance in the greater context of my goals and general interests in life, this effort represents far more to me than the mere sum of its results. In rudimentary terms this project has a set of requirements that may or may not have been accomplished in full this semester, but in many ways I like to think that this project abstractly represents the reason I am an engineer in the first place. I am a tireless advocate for the use of technology to further the cause of human progress and development, and I intend to spend my life engaged in the service of endeavors to this end. This project started for me long before the beginning of this particular semester. In many ways this is the culmination of several *years* of kicking ideas about solar technology and remote monitoring around in my head, but none of this would have ever begun had I not taken on the enormous challenge to go to The Gambia during the summer of 2007 to work on the construction of the Bobabo Memorial High School. For that opportunity I have to thank my good friend Bryan Reagan, who invited me to work with him on it in the first place. My colleagues in that undertaking, Jon Brown and Seku Barrow, were also sources of great support and insight.

Returning from that adventure, my closest friend at school, off whom I bounced this idea more times than I can count, was David Cusick. While he'd naturally refuse any notion of credit in this process, one of the biggest hurdles in my struggle to think about this project was the fact that I couldn't see the forest through the trees back in 2007 as far as the technology was concerned, and Dave, being a source of far greater knowledge than I, had the patience to discuss and tolerate my stupidity until I finally understood what was actually going on some years later. His role was subtle, but undeniable.

I also have to thank the many faculty members at the University of Florida that I've bothered with this project. Foremost among them is my faculty advisor and friend Dave Small, who, despite his yet unrevealed opinions about the viability of this idea to address the abstract and idealistic humanitarian goals that underlie my original intentions to pursue this project, supported my efforts nevertheless because he believed in me. Not only that, he is to be credited for most of the cool ideas along the way. I am incredibly lucky to have benefitted from Dave's utterly uncommon willingness to spend enormous amounts of time working with his students if they're just willing to *try*. Cheers! Also, Dr. Karl Gugel is rightly deserving of thanks for his advice on some of the hardware aspects of this project. It is a testimony to Dr. Gugel's integrity as a teacher that he would take all of the time that he did to entertain my ideas, especially considering the fact that I was not actually a student in any of his classes this semester! (One word—optoisolators!) Finally, I have to thank

Douglas Dankel, the last faculty member of my advisory panel, who came to the rescue and listened to this thesis at the last possible minute. His feedback was both constructive and insightful.

A lot of my fellow students are also deserving of mention and thanks, but in the interest of brevity I have to limit this list to my extremely resourceful friend Jose "Speedy" Morales, who managed without fail to have absolutely critical insights on some hardware issues of mine at precisely the moments I needed them. I don't attribute this to luck—that guy just knows his stuff.

Beyond this, I could not possibly consider these acknowledgements complete without the most wholehearted thanks to my entire family not only for their support of this project but also for their active engagement in my efforts. My sister Kristina is one of my most tireless advocates, and almost wholly to blame for encouraging me by example to spend endless sleepless nights working on this project. I must also thank my mom for her unyielding support in every major undertaking I have ever begun in life, despite the fact that she is no friend to computers. And, of course, I have to thank my dad, whose idea this entire project was in the first place. At one point during the semester prior to my departure for The Gambia, during a discussion of the costs involved with purchasing a customized cellular modem and paying some company a monthly fee for remote monitoring services, in a quote which stands out in my mind justifiably as the original genesis of this project, my dad burst out with an inspired remark of frustrated indignation, "You know you could just *build* this damn system yourself and it'd be cheaper!" This statement occurred at the end of an especially busy Spring Break spent constructing a 250 watt solar array on the roof of our house, not as a business venture or a clever way to screw FPL by running our pool pump off the sun, but as an old-fashioned science project—a pursuit of knowledge for knowledge's sake. (This system ended up being critical to the success of this project.) My dad is my greatest supporter and best friend, and the few good ideas I have lying around in my head are attributable to him.

I also would have liked to thank my late grandfather, who took such a personal interest in my efforts in school and my work abroad. The last two questions he ever asked me are burned in my memory and deserve repetition.

1.) "How is the school in Gambia doing?"
2.) "How are you doing in school?"

I hope this project serves, in some way, as an answer to both of these questions.